\documentclass[12pt]{article}

\usepackage{amsfonts,amsthm,amsmath,amssymb,upgreek,bm}
\usepackage[colorlinks=true, linkcolor=blue, citecolor=cyan, urlcolor=cyan]{hyperref}
\usepackage[paper=letterpaper,margin=.85in]{geometry}

\usepackage{mathrsfs}
\usepackage{float}
\usepackage{graphicx}
\usepackage[mathscr]{eucal}
\usepackage[T1]{fontenc}
\usepackage{pifont}
\usepackage{enumitem}
\usepackage{color}
\usepackage[dvipsnames]{xcolor}
\usepackage{bbold}
\usepackage{physics}
\usepackage[numbers,sort&compress]{natbib}
\usepackage{subcaption}
\usepackage{braket}

\def\p{\partial}
\def\O{\mathcal{O}}
\def\D{\mathcal{D}}
\def\M{\mathcal{M}}
\def\<{\langle}
\def\>{\rangle}
\def\R{\mathbb{R}}
\def\Z{\mathbb{Z}}

\def\ii{\mathrm{i}}
\numberwithin{equation}{section}

\begin{document}
\thispagestyle{empty}

\vspace*{2.5cm}
\begin{center}
{\bf {\LARGE A note on the quantization of \\ \vspace{10pt} angular momentum for black holes}}\\

\begin{center}
\vspace{1cm}

\vspace{1cm}
 {\bf Jon Goker${}^1$, Luca V. Iliesiu${}^1$,
Elisa Tabor${}^2$
}
\\
\bigskip \rm
\bigskip
${}^1$Leinweber Institute for Theoretical Physics and Department of Physics, University of California, Berkeley, California 94720, USA
\\
${}^2$ Leinweber Institute for Theoretical Physics, Stanford University, Stanford, CA 94305, USA
\rm
\end{center}
\rm
\end{center}

\vspace{1.5cm}
{\bf Abstract.}
{We argue that the gravitational path integral for rotating black holes is periodic in the angular velocity, implying the quantization of angular momentum in arbitrary dimensions for either asymptotically flat or AdS boundary conditions. In AdS$_3$, this periodicity is a consequence of the boundary mapping class group. In higher dimensions, the periodicity arises from an infinite family of saddles labeled by integer shifts of the angular velocity unrelated to the boundary mapping class group; summing over these saddles enforces quantization independently of any large boundary diffeomorphism. We construct these saddles explicitly for Kerr-Newman black holes in both asymptotically flat space and AdS$_4$, and observe that even the path integral for the 4D Schwarzschild black hole, typically the simplest case, receives contributions from an infinite set of rotating saddles.} 

\begin{quotation}
\noindent

\end{quotation}

\setcounter{page}{0}
\setcounter{tocdepth}{2}
\setcounter{footnote}{0}
\newpage

\parskip 0.1in

\setcounter{page}{2}
\setcounter{tocdepth}{2}
\tableofcontents

\section{Introduction}

The gravitational path integral provides a concrete mechanism for treating black holes as quantum systems, by interpreting the sum over gravitational saddles as a trace over an underlying Hilbert space \cite{Gibbons:1976ue}. Whereas for an ordinary quantum system, the discreteness of quantum numbers is unsurprising, in gravity it is a nontrivial consequence of the sum over geometries. In particular, in classical general relativity, the angular momentum of a black hole is defined as an asymptotic conserved charge \cite{Brown:1992br, Brown:1992bq} and does not manifestly carry a discrete representation label. The goal of this note is to argue that the gravitational path integral is always periodic in the angular potential, which in turn implies the quantization of angular momentum for rotating black holes in arbitrary dimensions, with either asymptotically flat or AdS boundary conditions.

In AdS$_3$, the quantization of angular momentum for BTZ black holes has been shown explicitly using the gravitational path integral
\cite{Maldacena:1998bw, Maloney:2007ud}. There, the path integral is computed with torus boundary conditions; fixing the modulus of the torus sets the temperature and angular velocity. An infinite family of saddles then contributes: each is obtained from the standard BTZ geometry, in which the thermal circle is contractible, by a boundary mapping class group ($SL(2,\Z)$) transformation that selects a different contractible cycle. Summing over these geometries produces a path integral that is invariant under periodic shifts of the angular potential, and Fourier transforming to the canonical ensemble then yields a discrete spectrum for the BTZ angular momentum. If black holes in higher dimensions are to be treated as quantum systems, it is natural to ask what mechanism plays the role of the sum over BTZ geometries in higher dimensions. 

In this paper, we will show that, just as in AdS$_3$, when the angular velocity or angular momentum of a black hole is fixed, there is an infinite family of black hole saddles whose sum is periodic in the angular potential and enforces the quantization of angular momentum.\footnote{Similar arguments were discussed for the quantization of electric charge in black holes \cite{Coleman:1991ku, Kolanowski:2026gii} and for the quantization of angular momentum for slowly rotating black holes close to extremality \cite{Iliesiu:2020qvm, Heydeman:2020hhw, Chen:2023mbc}. These arguments were inspired by the sum over saddles encountered in the super-Schwarzian theory in \cite{Stanford:2017thb}. A similar sum over saddles was also discussed in the context of supersymmetric indices for black holes \cite{Cabo-Bizet:2018ehj, Cabo-Bizet:2025kjc, Cassani:2019mms, Bobev:2019zmz, Bobev:2020pjk, Larsen:2021wnu, Hristov:2021qsw, Hristov:2022pmo, BenettiGenolini:2023rkq, Iliesiu:2021are, H:2023qko, Anupam:2023yns, Boruch:2023gfn, Hegde:2023jmp, Chowdhury:2024ngg, Chen:2024gmc, Cassani:2024kjn, Hegde:2024bmb, Adhikari:2024zif, Heydeman:2024fgk, Boruch:2025qdq, Cassani:2025sim, BenettiGenolini:2025jwe, Bandyopadhyay:2025jbc, Iliesiu:2022kny} and rotating supersymmetric black holes \cite{Boruch:2022tno}; however, in that setting, most saddles (with the exception of two saddles that admit Killing spinors) have a vanishing contribution to the path integral due to fermionic zero modes being present on such backgrounds. }
While the concrete examples we study are in three and four spacetime dimensions, the mechanism is general: for any solution of general relativity with a horizon, in any spacetime dimension, and for either asymptotically flat or AdS boundary conditions, there always exists an infinite family of saddles whose sum enforces the quantization of angular momentum. Perhaps surprisingly, even the path integral for the simplest black hole, the 4D Schwarzschild black hole, for which the asymptotic angular velocity vanishes, receives contributions from a family of rotating saddles.

The rest of this paper is structured as follows. In Section \ref{section: maxwell}, we start with a pedagogical example of Maxwell theory from the saddlepoint approximation of the path integral, explaining how charge quantization appears from an infinite sum over saddles. Section \ref{section: brown-york} reviews the Brown–York formalism and explains how different choices of fixed boundary data correspond to distinct gravitational ensembles. Section \ref{sec: 3d} constructs explicit fixed angular-velocity and fixed angular-momentum saddles in 3D gravity, the sum of which enforces the half-integer spectrum of the angular momentum $J_z$ around some axis of rotation labeled by $z$. Section \ref{sec: 4d} and Appendix \ref{sec:AdS4} extend the analysis to 4D, in asymptotically flat and asymptotically AdS$_4$ spacetimes, respectively. For both cases, we present a new family of smooth saddles in the grand canonical and canonical ensembles. In Appendix \ref{app: convergence}, we analyze the convergence of the sum over the family of rotating saddles in all previously studied cases. Finally, in Appendix \ref{app:variance}, we discuss how the sum over saddles is particularly important in the near-extremal limit, where we show that even the average value of the angular momentum gets arbitrarily close to a half-integer.  

\section{A pedagogical example: \\ 
charge quantization from a gauge theory path integral} \label{section: maxwell}
As a pedagogical example that shares numerous similarities with the quantization of black hole angular momentum, we first review the path integral quantization of charge in $2D$ Maxwell theory. To see how charge is quantized, we will compute the disk path integral in this theory with different choices of boundary conditions, fixing first the chemical potential and later the charge.

To start, we compute the disk partition function with fixed chemical potential, for which we impose Dirichlet boundary conditions on the gauge field. The Euclidean action takes the form
\begin{equation} \label{eq:Maxwell Action Dirichlet}
    I(\beta,\mu_E) = \frac{1}{4e^2}\int_{D^2} \sqrt{g}F_{\mu \nu}F^{\mu \nu} \,,
\end{equation}
where $F=dA$ and the coupling constant of the vector potential is given by $e$. We fix the vector potential on the boundary, $\delta A \big|_{\partial D^2} = 0$. Taking the background metric to be flat and given by $ds^2=dr^2+r^2d\tau^2$, we shall work in a gauge in which $\partial_\tau A_r=0$. Using the equations of motion, the vector potential can be written as 
\begin{equation} \label{eq:AphiFixed}
    A_\tau = \left(\mu_E-\frac{2\pi n}{\beta}\right) \frac{r^2}{r_*^2} + \frac{2\pi n}{\beta} \,, \quad n \in \mathbb{Z} \,,
\end{equation}
    where $r_*$ denotes the boundary value of $r$ and where $\mu_E$ sets the boundary value of $A_\tau|_{r=r_*} = \mu_E$, having the interpretation of a chemical potential. This fixes the holonomy of $A$ along the boundary thermal circle to $e^{\ii \oint_{r=r_*} A } = e^{\ii\mu_E\beta}$. Requiring the field strength to be smooth fixes the holonomy of $A$ near the origin to be trivial, $e^{\ii\oint A} = 1$.\footnote{Note that in configurations where $\oint A \neq 0$ around the origin, the gauge field has to be specified in patches, with one patch including the origin and another excluding the origin. The condition needed to be able to glue these two patches is precisely that $e^{\ii\oint A} = 1$ around the origin.} Each integer $n$ labels a distinct vector potential, with configurations with different $n$ not related by gauge transformations.
Evaluated on-shell, the action simplifies to
\begin{equation}
    I_{\text{cl}}(\beta,\mu_E,n) = \frac{\beta}{e^2r_*^2}\left(\mu_E-\frac{2\pi n}{\beta}\right)^2 \,, \label{eq:OnShellAction}
\end{equation}
and the partition function in the saddlepoint approximation sums over contributions from each $n\in\Z$,
\begin{equation}
    Z(\beta,\mu_E) = \int \D A_\tau e^{-I[A_\tau]} = \frac{2\sqrt\pi}{er_*\sqrt\beta} \sum_{n=-\infty}^{\infty} \exp\left(\frac{-\beta}{e^2r_*^2}\left(\mu_E-\frac{2\pi n}{\beta}\right)^2\right) \,.
\end{equation}
Above, we have also explicitly included the one-loop determinant around each saddle \cite{Witten:1991we, Witten:1992xu, Cordes:1994fc}.
The partition function is periodic under $\mu_E \sim \mu_E +\frac{2\pi}\beta$. This periodicity in $\mu_E$ allows us to perform a Poisson resummation,
\begin{align}
    Z(\beta,\mu_E) &= \sum_{Q\in\Z} e^{\ii\beta\mu_E Q} Z(\beta, Q) = \sum_{Q\in\Z} e^{\ii\beta\mu_E Q} \frac{\beta}{2\pi} \int_{-\infty}^{\infty} d\mu' e^{-\ii\beta\mu' Q} \frac{2\sqrt\pi}{er_*\sqrt\beta} e^{\frac{-\beta}{e^2r_*^2}(\mu')^2}\nonumber \\
    &= \sum_{Q\in\Z} e^{\ii\beta\mu_E Q} e^{-\frac14\beta (er_*Q)^2} \label{eq:gc Z}
\end{align}
where we see that the Poisson dual variable to the chemical potential is the quantized charge $Q$. The last equation is the conventional expression for the partition function of Maxwell theory on a disk obtained from canonical quantization \cite{Migdal:1975zg, Witten:1991we, Blau:1991mp}. What we learn from this pedagogical example is that to have a disk partition function that can be expressed as a sum over integer charges, one needs an infinite number of saddles whose sum is necessary in order to make the partition function periodic in the boundary value of the gauge field that sets the chemical potential. 

Note that while the charge $Q$ in \eqref{eq:gc Z} is quantized, the semiclassical charge associated with the saddles with different $n$ is not. The charge within each saddle can be defined as
\begin{align}
    (Q_\text{cl})_n = -\frac{1}{i\beta} \partial_{\mu_E} I_{\text{cl}}(\beta, \mu_E,n)\,,
    \label{eq:charge-in-each-saddle}
\end{align} 
which is clearly not quantized. However, note that the average value of $Q_\text{cl}$ is the same as the average of $Q$ in \eqref{eq:gc Z}, 
\begin{align}
    \langle Q \rangle =  \frac{1}{Z(\beta, \mu_E)} \sum_{Q\in\Z} Q \,e^{\ii\beta\mu_E Q} e^{-\frac14\beta (er_*Q)^2}  = \frac{1}{Z(\beta, \mu_E)}\frac{2\sqrt\pi}{er_*\sqrt\beta} \sum_{n=-\infty}^{\infty} (Q_\text{cl})_n \exp\left(\frac{-\beta}{e^2r_*^2}\left(\mu_E-\frac{2\pi n}{\beta}\right)^2\right) \,.
    \label{eq:average-of-Q}
\end{align}
The same equality applies to higher moments of the charge, as $\langle Q^n \rangle$ is the same as the one obtained by taking the $n$-th derivative of the action of each saddle, generalizing \eqref{eq:charge-in-each-saddle} and \eqref{eq:average-of-Q}.

As an alternative to the above quantization procedure, we can construct the partition function in the charge microcanonical ensemble, in which we impose Neumann boundary conditions on the action. Fixing the charge amounts to fixing the field strength on the boundary, $F_{r\tau}\big|_{\partial D^2} = \ii Qe^2r_*$. To have a well-defined variational principle, we need to add an extra boundary term to Eq.~\eqref{eq:Maxwell Action Dirichlet},
\begin{equation}
    I(\beta,Q) = \frac{1}{4e^2}\int_{D^2} \sqrt{g}F_{\mu \nu}F^{\mu \nu} + \frac{1}{e^2}\int_{\partial D^2} \sqrt{h}A_\mu F^{\mu\nu}n_\nu + I_{\rm ct}\,,
\end{equation}
where $I_{\rm ct}$ is a constant counterterm that will become relevant later on. After imposing the boundary conditions as well as the smoothness condition on the holonomy of $A$ near the origin, we find the vector potential 
\begin{equation}
     A_\tau = \frac{\ii Qe^2}{2}r^2 + \frac{2\pi n}{\beta}\,, \quad n \in \mathbb{Z} \,,
\end{equation}
which are related by large gauge transformations $\lambda=\frac{2\pi n}{\beta}\tau$ that correspond to windings around the disk. Since these change the boundary data, they contribute physically distinct solutions to the partition function, as can be seen from the fact that the on-shell action of the different solutions is $n$-dependent, 
\begin{equation}
    I_{\text{cl}}(\beta,Q,n) = \frac{\beta (Qer_*)^2}{4} - 2\pi\ii Qn \,.
\end{equation}
By summing over this family of saddles and including the counterterm $I_{\rm ct}$, we find the microcanonical partition function
\begin{equation}
    Z(\beta, Q) = e^{-\frac14\beta (er_*Q)^2} e^{-I_{\rm ct}} \sum_{n \in\Z} e^{2\pi\ii Qn} \,.
\end{equation}
To make sense of the sum, we introduce a small regulator $\epsilon>0$, with which
\begin{equation}
    \sum_{n \in\Z} e^{2\pi\ii Qn} = \sum_{n\in\Z} e^{-\epsilon n^2} e^{2\pi\ii Qn}=\sqrt\frac{\pi}{\epsilon} \sum_{m\in\Z} e^{-\frac{\pi^2}{\epsilon}(m-Q)^2}
\end{equation}
after Poisson resummation. Then let $I_{\rm ct} = -\frac12\log\frac{\epsilon}{\pi}$, such that
\begin{equation}
    Z(\beta, Q) = \left\{ \begin{matrix} e^{-\frac14\beta (er_*Q)^2} & Q\in\Z \\ 0 & \text{else} \end{matrix} \right. \,, \label{eq:mc Z}
\end{equation}
resulting in the quantization of charge from a different perspective. Once again, an infinite number of saddles is required for us to observe the quantization of the charge. Whereas in the previous case with Dirichlet boundary conditions and fixed chemical potential, the quantization originated in the bulk term, for the fixed charge ensemble, the quantization came purely from the addition of the boundary term.

In the following sections, we use similar methods to probe the statistical mechanics of charged rotating black hole solutions in three and four dimensions.

\section{Brown-York formalism} 
\label{section: brown-york}
In gravitational theories, the mechanism of angular momentum quantization directly parallels the charge quantization in Maxwell theory. Just as the smoothness of the field strength at the origin imposes the periodicity of the boundary holonomy, we will see that the smoothness at the Euclidean horizon imposes periodicity of the angular potential, and just as summing over Maxwell saddles corresponding to shifts of the holonomy quantizes the conjugate electric charge, we will see that summing over gravitational saddles corresponding to shifts of the angular velocity quantizes the conjugate angular momentum. In preparation for generalizing our discussion to higher dimensions and other ensembles, in this section, we first discuss how to fix either angular velocity or angular momentum in gravity. We do so by reviewing the Brown-York formalism \cite{Brown:1992br, Brown:1992bq}.

Brown and York \cite{Brown:1992br, Brown:1992bq} reframed the gravitational path integral as a functional of the induced metric $h_{ij}$ on a timelike boundary and defined the quasi-local holographic stress tensor,
\begin{equation} \label{eq:brownyork}
    T_{ij} = \frac{2}{\sqrt{-h}}\frac{\delta I}{\delta h^{ij}} \,,
\end{equation}
conjugate to metric deformations. This object packages the boundary energy and momentum densities, allowing us to pass between thermodynamic ensembles via a Legendre transform at the boundary. Below, we review how fixing the grand canonical ensemble $(\beta,\Omega)$ amounts to fixing the boundary metric, while fixing the canonical ensemble $(\beta,J)$ amounts to fixing a combination of components of the boundary metric and components of $T_{ij}$.\footnote{See also \cite{Hawking:1995fd, Booth:1998eh, Regge:1974zd, Szabados:2009eka} for related applications of the Hamiltonian formalism, \cite{Caldarelli:1999xj, Dolan:2014mra} for comparisons of fixed angular velocity and momentum ensembles, and \cite{Compere:2008us, Krishnan:2016mcj, Krishnan:2016tqj, Marolf:2018ldl, Chua:2023srl} for other related works on mixed/Neumann boundary conditions in gravity. An analogous discussion for the ensembles of charged black holes was initiated in \cite{Braden:1990hw, Hawking:1995ap}.}

\subsection{Fixed angular velocity}
We consider the Euclidean Einstein-Hilbert action in arbitrary dimension in the grand canonical ensemble, in which we fix the inverse temperature $\beta$ and angular velocity $\Omega_0$,
\begin{equation}
    I(\beta,\Omega_0) = -\frac{1}{16\pi G_N} \int_{\M} d^D x\sqrt{g}(R-2\Lambda) + \frac{1}{8\pi G_N}\int_{\p\M}d^{D-1}x \sqrt{h} K \,,
\end{equation}
where $G_N$ is Newton's constant in $D$ dimensions and $\Lambda$ is the cosmological constant. This formalism can also be applied to the Einstein-Maxwell action, which we shall use when analyzing Kerr-Newman saddles later in this paper. We choose coordinates
$$x^\mu=(\tau, r, \theta^A,\phi)\,,\quad A=1,\ldots,D-3$$
where $r$ is an asymptotic radial coordinate and $(\theta^A,\phi)$ denote the angular coordinates. We will consider axisymmetric black hole saddles that rotate along the axis transverse to $\phi$, given by
\begin{equation}
    ds^2 = N^2d\tau^2 + h_{ij}(dx^i-\ii V^id\tau)(dx^j-\ii V^jd\tau)\,,\quad i=2,\ldots,D\,,
    \label{eq:decomposition-metric}
\end{equation}
with boundary topology $S^1\times S^{D-2}$ and asymptotic metric\footnote{Note that $\Omega_0 \in \R$ is an angular velocity that is obtained from a Wick rotation of a Lorentzian rotating solution. In Euclidean signature, a real metric corresponds to $\Omega_0 \equiv -\ii \Omega_e
\in -\ii \R$.}
\begin{equation}
    ds^2 = \eta_{\alpha\beta}dx^\alpha dx^\beta + \eta_{\phi\phi} (d\phi' - \ii\Omega_0 d\tau)^2 \,.
\end{equation}
Above, the indices $\alpha,\beta=1,\dots,D-1$ run over all coordinates except $\phi$. In order to fix the angular velocity at the boundary, we fix the leading asymptotic component of $V^\phi$ as follows
\begin{equation}
    \lim_{r\to\infty} V^\phi \equiv \Omega_0 \,.
\end{equation}
We can think of this as imposing Dirichlet boundary conditions on the path integral.

The energy surface density and momentum surface density are defined from the Brown-York stress tensor $T^{ij}$ as \cite{Brown:1992br}
\begin{align}
    \epsilon &\equiv u_iu_j T^{ij} = -\frac{1}{\sqrt{\sigma}}\frac{\delta I_{\text{cl}}}{\delta N}, \\
    j_a &\equiv -\sigma_{ai}u_jT^{ij} = \frac{1}{\sqrt{\sigma}}\frac{\delta I_{\text{cl}}}{\delta V^a} \,,
\end{align}
where $I_{\text{cl}}$ is the classical on-shell action, computed by solving the Einstein equations of motion, and $\sigma$ is the metric for a boundary Cauchy slice $\Sigma$. We now rewrite the GHY boundary term as
\begin{equation} \label{eq:hamiltonianbdy}
    \frac{1}{8\pi G_N}\int_{\p\M} d^{D-1}x \sqrt{h}K = \int_{\p\M}d^{D-1}x \left(\sqrt{\sigma}\left[N\epsilon-V^aj_a\right] - \frac{1}{8\pi G_N} \sqrt{h}t_\mu K^{\mu\nu}\p_\nu t\right) \,,
\end{equation}
where $t$ is a scalar field on the boundary $\p\M$ that labels the foliation on which $\Sigma$ is defined, and $t_\mu$ is a time vector field on $\p\M$ that specifies the time direction \cite{Brown:1992bq}. The mass and angular momentum (about a given axis that we call $z$) associated with each classical saddle are given by
\begin{align}
    M = \int_{\Sigma} d^{D-2}x\sqrt{\sigma} \epsilon \,,\quad J_{z,\text{cl}} = \int_{\Sigma} d^{D-2}x\sqrt{\sigma} j_\phi \,.
\end{align}
We thus find that the terms in \eqref{eq:hamiltonianbdy} can be rewritten as
\begin{equation}
    \int_{\p\M}d^{D-1}x \sqrt{\sigma}N\epsilon = \beta M \,,\quad \int_{\p\M}d^{D-1}x\sqrt{\sigma}V^aj_a = \beta\Omega_0 J_{z,\text{cl}}
\end{equation}
while for the tilt term,
\begin{align}
    \frac{1}{8\pi G_N} \int_{B_H}d^{D-1}x\sqrt{h} t_\mu K^{\mu\nu}\p_\nu t &= -\frac{1}{8\pi G_N} \int_{B_H}d^{D-1}x\sqrt{h} (Nu_\mu+V_\mu) K^{\mu\nu} \frac1N u_\nu \nonumber\\
    &= \frac{1}{8\pi G_N} \int_{B_H}d^{D-1}x\sqrt{h} n^i\p_i\log N = \frac{1}{8\pi G_N} \int_0^\beta d\tau \int d^{D-2}x\sqrt{\sigma} \frac{2\pi}{\beta} \nonumber\\
    &= \frac{A_H}{4 G_N} = S \,,
\end{align}
where the integral is evaluated over the black hole horizon. We conclude that in the grand canonical ensemble, the classical on-shell action takes the form\footnote{In 4D for fixed charge $Q$, this form of the on-shell action can be related to the more typical form in \cite{Gibbons:1976ue} via the Smarr formula $M=2TS+2\Omega J_z+Q\Phi$ \cite{Smarr:1972kt}.}
\begin{equation} \label{eq:onshell fixed Omega}
    -I_{\text{cl}} (\beta,\Omega_0) = -\beta M + \beta\Omega_0J_{z,\text{cl}} +S \,,
\end{equation}
where $M$, $J_{z,\text{cl}}$, and $S$ are functions of $\beta$ and $\Omega_0$.

\subsubsection*{Quantization of angular momentum with Dirichlet boundary conditions}

We will show explicitly that in three- and four-dimensional gravity, for a fixed angular velocity, imposing a smooth geometry at the black hole horizon leads to a family of solutions, each with distinct angular momenta.\footnote{ The angular momenta are given by \eqref{eq:JM_n} for 3D BTZ, \eqref{eq:JMn_4d} for 4D Kerr-Newman, and \eqref{eq:JMn_AdS4} for AdS$_4$ Kerr-Newman.} The gravitational path integral with fixed angular velocity $\Omega_0$ (i.e., fixed $V^{\phi}|_{r\to \infty}$ in the Brown-York formalism) sums over these saddles, where each on-shell action depends on the angular velocity $\Omega_0$ through the integer shifted value $\Omega_0+\frac{4\pi\ii}{\beta}\mathbb Z$,
\begin{equation} \label{eq: general Dirichlet quantization}
\begin{split}
    Z_{\text{grav}}(\beta,\Omega_0) &= \sum_{n\in\Z} e^{-I_{\text{cl}}\left(\beta,\Omega_0+\frac{4\pi\ii}{\beta}n\right)} Z_{\text{1-loop}}\left(\beta,\Omega_0+\frac{4\pi\ii}{\beta}n\right) (1+\ldots) \,.
\end{split}
\end{equation}
At the level of the on-shell action, we see that the resulting partition function is periodic in $\Omega_0$. Since $J_z$ is canonically conjugate to the now-periodic angular velocity $\Omega_0$, we hence see that $J_z$ is quantized. However, as with the classical value of the charge $(Q_\text{cl})_n$ in Maxwell theory, the individual Brown-York angular momenta $(J_{z,\text{cl}})_n$  are not quantized.

The one-loop determinant and further subleading corrections have the same $n$-dependence, and hence the same periodicity in $\Omega_0$, as the on-shell action. These corrections can be understood as fluctuations about each saddle, where the parameter $n$ labels the saddle. In \eqref{eq: general Dirichlet quantization} and in later sections, we implicitly assume that all such saddles contribute. This might not necessarily be the case, as not all saddles need to be picked up on a given integration contour of the gravity path integral.\footnote{For instance, not all saddles satisfy the proposed KSW criterion for which saddles contribute to the path integral \cite{Witten:2021nzp}.} While answering which saddles contribute to the GPI is generally difficult, there are cases where we know that not all saddles in \eqref{eq: general Dirichlet quantization} contribute: for instance, the sum over $n$ can sometimes be divergent,\footnote{We analyze such cases in Appendix \ref{app: convergence}. See also \cite{Witten:2021nzp, Mahajan:2025bzo, Singhi:2025rfy, Kolanowski:2026gii} for related contour/convergence issues for complex black hole saddles.} whereas the sum over angular momenta is provably convergent. Agreement between the two ensembles thus requires summing over only a finite number of saddles in \eqref{eq: general Dirichlet quantization}.\footnote{A similar phenomenon was encountered for the quantization of black hole electric charges \cite{Coleman:1991ku}.} Even if only finitely many saddles contribute, the partition function is nevertheless periodic in $\Omega_0$, still guaranteeing the quantization of the angular momentum.

Finally, we note importantly that the quantization of angular momentum is independent of the choice of the rotation axis (for $D>3$). For example, in $D=4$, we are free to introduce an arbitrary choice of rotation axis, $(\theta_{\hat{n}}, \phi_{\hat{n}})$, with the corresponding Kerr-Newman solution related by diffeomorphism to the original. With this reparameterization, we repeat the analysis above, finding that the periodicity in $\Omega_i$ is independent of the choice of axis of rotation and analogously leads to a quantized $J_i$ by Fourier transform. With this, we can construct solutions for rotations about all three coordinate axes with quantized $J_x, J_y,$ and $J_z$ respectively. These operators are the generators of the full $\mathfrak{su}(2)$ Lie algebra, and our partition function runs over the Cartan subgroup, $U(1)$. Because the partition function in \eqref{eq: general Dirichlet quantization} is independent of the axis of rotation, the partition function $Z(\beta, \Omega_0)$ is a class function over $SU(2)$ that is completely determined by its values on the Cartan subgroup parameterized by $g=e^{\Omega_0\beta J_{\hat n} }$. Any class function admits a decomposition into a sum over characters of the group,
\begin{equation}
    Z(\beta, \Omega_0) = \sum_{j \in \frac{1}{2}\mathbb{Z}^+} \chi_j(\ii\Omega_0 \beta)Z_j(\beta)
\end{equation}
where the sum over $j \in \frac{1}{2}\mathbb{Z}^+$ shows that the Hilbert space arranges itself into $SU(2)$ irreducible representations. The above argument directly generalizes to arbitrary $D$.

\subsection{Fixed angular momentum}
We now turn to Euclidean Einstein-Hilbert in the canonical ensemble, in which we fix the inverse temperature $\beta$ and angular momentum $J_z$. We will focus on showing the quantization for an individual component of the angular momentum, though the argument discussed above can be used to show that the angular momentum falls in representations of the appropriate rotation group. Instead of fixing the leading in $r$ component of $V^\phi$, we want to fix the momentum surface density $j_\phi$ (that determines $J_z = \int_{\Sigma} d^{D-2}x\sqrt{\sigma} j_\phi$). Incorporating the appropriate boundary term and counterterm leads to the action
\begin{equation} \label{eq: Brown-York bdy term}
    I(\beta,J_z) = I(\beta,\Omega_0) + \int_{\p\M} d^{D-1}x \sqrt{\sigma} V^aj_a + I_{\rm ct} \,.
\end{equation}
In addition to fixing $j_\phi$, we impose the boundary topology $S^1\times S^{D-2}$ with asymptotic metric
\begin{equation}
    ds^2 = \eta_{\alpha\beta}dx^\alpha dx^\beta + \eta_{\phi\phi} (d\phi' - \ii\Omega d\tau)^2 \,,
\end{equation}
where the indices $\alpha,\beta=1,\dots,D-1$ again run over all coordinates except $\phi$, and where
\begin{equation} \label{eq:Omega_n from V^phi}
    \Omega \equiv \lim_{r\to\infty} V^\phi,
\end{equation}
is now a free parameter. We can think of this as imposing Neumann boundary conditions on the path integral.

We conclude that in the canonical ensemble, the classical on-shell action takes the form\footnote{In the microcanonical ensemble, we would have
\begin{equation}
    I(E,J_z) = I(\beta,\Omega) - \int_{\p\M}d^{D-1} x\sqrt{\sigma} (N\epsilon - V^aj_a)\Big|_{\text{on-shell}} = - S\,.
\end{equation}
On the other hand, if we want to pass to the grand canonical ensemble, in which the chemical potential for charge is fixed, we would have 
\begin{equation}
    I(\beta,\Omega,\Phi) = I(\beta,\Omega,Q) - \int_{\p\M}d^{D-1}x\sqrt{\sigma} \Phi\rho\Big|_{\text{on-shell}} = \beta M - \beta\Omega J_{z,\text{cl}} - \beta\Phi Q - S\,,
\end{equation}
where $\Phi\equiv -u^\mu A_\mu\Big|_{\Sigma_\infty}$, $\rho\equiv \frac{1}{e^2}n_\mu F^{\mu\nu}u_\nu$, and $Q=\int_{\Sigma_\infty}\sqrt{\sigma}\rho$ is the total electric charge.}
\begin{equation} \label{eq:onshell canon}
    -I_{\text{cl}}(\beta,J_z) = -\beta M + S \,.
\end{equation}

\subsubsection*{Quantization of angular momentum with Neumann boundary conditions}
We will show explicitly in three and four-dimensional gravity that for a fixed angular momentum, imposing a smooth geometry at the black hole horizon leads to a family of solutions parametrized by different angular velocities $\Omega_n$, which are given by \eqref{eq:3D neumann velocities} for 3D BTZ, \eqref{eq:4d Neumann velocities} for 4D Kerr-Newman, and \eqref{eq:ads4 Neumann velocities} for AdS$_4$ Kerr-Newman. We compute the fixed $J_z$ partition function from a Fourier transform of the fixed $\Omega$ partition function via a saddlepoint approximation: 
\begin{equation} \label{eq:brownyork fixedJ}
\begin{split}
    Z_{\text{grav}}(\beta,J_z) &= \frac{1}{4\pi} \int_{-\infty}^{\infty} d\theta e^{-\ii\theta J_z} Z(\beta,\theta) \\
    &\approx \sum_{n\in\Z} e^{-\beta J_z(\Omega_n-\Omega_0)-\beta M+S} = e^{S-\beta M} \sum_{n\in\Z} e^{4\pi\ii n J_z}
\end{split}
\end{equation}
where the saddlepoint approximation results in a sum over saddles $\theta=-\ii\beta\Omega_n$, each of which has the same $\beta$ and $J_z$. Just as in Maxwell theory, the $n$ dependence is fully contained in the new boundary term,
\begin{equation}
    -I_{\text{cl}}(\beta,J_z) = -I_{\text{cl}}(\beta,\Omega_0) - \beta\Omega_n J_z = -\beta M + S + 4\pi\ii n J_z\,.
\end{equation}
The full gravitational path integral with fixed angular momentum $J_z$ sums over these classical saddles along with quantum corrections,
\begin{align}
    Z_{\text{grav}}(\beta,J_z) &= \sum_{n\in\Z} e^{-I_{\text{cl}}(\beta,J_z)} Z_{\text{1-loop}}(\beta,J_z) (1+\ldots) = e^{S - \beta M} Z_{\text{1-loop}}(\beta,J_z) (1+\ldots) \sum_{n\in\Z} e^{4\pi\ii n J_z}
\end{align}
where $e^{S - \beta M} Z_{\text{1-loop}}(\beta,J_z) (1+\ldots) $ includes the on-shell piece as well as subleading terms, which are all independent of $n$. Just as in the 2D Maxwell case, we introduce a regulator $\epsilon>0$, with which 
\begin{equation}
    \sum_{n \in\Z} e^{4\pi\ii nJ_z} = \sum_{n\in\Z} e^{-\epsilon n^2} e^{2\pi\ii n(2J_z)}=\sqrt\frac{\pi}{\epsilon} \sum_{m\in\Z} e^{-\frac{\pi^2}{\epsilon}(m-2J_z)^2}
\end{equation}
after Poisson resummation. We again let $I_{\rm ct} = -\frac12\log\frac{\epsilon}{\pi}$, such that
\begin{equation} \label{eq: general Neumann quantization}
    Z_{\text{grav}}(\beta, J_z) = \left\{ \begin{matrix} e^{S - \beta M} Z_{\text{1-loop}}(\beta,J_z) (1+\ldots)  & J_z\in\frac12\Z \\ 0 & \text{else} \end{matrix} \right. \,.
\end{equation}

\section{Angular momentum in 3D gravity} \label{sec: 3d}

We now turn to the pedagogical example of pure 3D gravity and explicitly write down the family of saddles that lead to the quantization of angular momentum, for both the fixed angular velocity and the fixed angular momentum ensembles. We will see that the analogue of the Maxwell smoothness condition at the origin is now smoothness at the horizon for our Euclidean saddles. 

\subsection*{3D saddles at fixed angular velocity}
We consider solutions to the fixed $\Omega_0$ Euclidean path integral in AdS$_3$ with Einstein-Hilbert action:\footnote{Throughout this section we let $\ell=1$.}
\begin{equation}
    I_{\text{EH}}(\beta,\Omega_0) =  -\frac{1}{16\pi G_N} \left[ \int_{\M_3} d^3x\, \sqrt{g^{(3)}} \left( R^{(3)} + 2 \right) + 2 \int_{\p\M_3} d^2x\, \sqrt{\gamma} \left( \kappa^{(3)} - 1 \right) \right] \,, \label{eq:BTZ Action}
\end{equation}
and impose the boundary topology of a torus $T^2$ with an asymptotic metric
\begin{equation}
    ds^2 = r^2dt_E^2 + \frac{dr^2}{r^2} + r^2(d\phi' - \ii\Omega_0 dt_E)^2 + \ldots\,,\quad r\to\infty \,,
\end{equation}
where we fix the Euclidean angular velocity $\Omega_0$ at the boundary and the corotating coordinate $\phi'$ is related to the usual BTZ coordinate $\phi$ via the diffeomorphism 
\begin{equation} \label{eq:corotating_diffeo}
    \phi' = \phi + \ii\Omega_0t_E\,.
\end{equation}
Since \eqref{eq:corotating_diffeo} is merely a coordinate change, to fix the angular velocity, we also have to describe the periodicity condition when going around the thermal boundary for probe fields that carry spin. For this, consider a fermionic field propagating on this background, with periodicity given by
\begin{equation}
    \psi(t_E, r, \phi) = -e^{\frac{\Omega_0\beta}2}\psi(t_E+\beta, r, \phi) = -\psi(t_E, r, \phi+2\pi)\,.
\end{equation}
Using the coordinate $\phi'$, this becomes
\begin{equation} \label{eq:fermionBC3d}
    \psi(t_E, r, \phi') = -\psi(t_E+\beta, r, \phi') = -\psi(t_E+\beta, r, \phi'+4\pi)
\end{equation}
where the second equality accounts for the fact that fermions are antiperiodic around $\phi\sim\phi+2\pi$.\footnote{If we consider a theory without any fermionic degrees of freedom, then the spacetimes that we sum over do not necessarily need to have a valid spin structure. In this case, we can describe the periodicity conditions that fix the angular velocity for a bosonic spinning field.\label{bosonic-vs-fermionic}}

The finite-action classical solutions of 3D gravity are fully classified and referred to as the PSL$(2,\Z)$ black holes \cite{Maldacena:1998bw, Maloney:2007ud}, see \cite{Maxfield:2020ale} for a nice review. On manifolds that admit fermionic fields, the gravitational path integral requires the additional input of a fixed boundary spin structure \cite{Boruch:2026hbr}. Here we will focus on the torus spin structure in which fermions are antiperiodic along both cycles, resulting in a partition function over the NS Hilbert space consistent with \eqref{eq:fermionBC3d}. Since the modular group must not only preserve the torus, but also its spin structure, this partition function is only invariant under the subgroup $\Gamma_\theta$ of PSL$(2,\Z)$, defined by
\begin{equation}
    \Gamma_\theta \equiv \left\{ \begin{pmatrix}
        a&b \\ c&d
    \end{pmatrix} \in\text{PSL}(2,\Z) \, | \, a+b = c + d = 1 \mod2 \right\} \,.
\end{equation}
The above boundary conditions fix the modular parameter of the boundary torus, $\tau = \frac{\ii}{2\pi}(\beta\Omega_0+\beta)$. The modular transforms $\gamma\in \Gamma_\theta$ are large diffeomorphisms that act on $\tau$. The family of on-shell saddles that fill these boundary conditions is given by the sum over modular images of thermal AdS,
\begin{equation} \label{eq:Gtheta}
    Z_{\rm 3D}(\tau,\bar{\tau}) = \sum_{\gamma\in\Z \backslash\Gamma_\theta} Z_{\text{TAdS}}(\gamma\tau,\gamma\bar\tau)\,,
\end{equation}
where $\gamma\in\Z \backslash\Gamma_\theta$ identifies $\gamma\sim T^{2n}\gamma$ for $n\in\Z$ \cite{Boruch:2026hbr}. Here we will proceed in a manner that generalizes to higher dimensions by deriving the family of shifted saddles for the BTZ saddle, given by an $S$-transform of thermal AdS. We will later show that these shifted saddles correspond to a subset of $\Gamma_\theta$ given by $ST^{2n}$ with $n\in\Z$.

The BTZ solution \cite{Banados:1992wn} takes the form 
\begin{equation}
    ds^2 = \frac{\Delta(r)}{r^2}\, dt_E^2 + \frac{r^2 dr^2}{\Delta(r)} + r^2 \left[d\phi' + \ii\left(\frac{r_+ r_-}{r^2} - \Omega_0\right) dt_E \right]^2\,, \quad \Delta(r) = (r^2 - r_+^2)(r^2 - r_-^2) \,, \label{eq:rotatingBTZMetric}
\end{equation}
where $r_+$ is positive and $r_-$ is imaginary in Euclidean signature. Note that the coefficient of $r^2 dt_Ed\phi$ determines the angular velocity, while the subleading in $r$ coefficient of $dt_Ed\phi$ determines the angular momentum 
\begin{equation} \label{eq:BTZ J}
    J_{z,\text{cl}}=\frac{1}{4G_N}r_+r_- \,.
\end{equation}

In this ensemble, all the saddles share the same fixed $\Omega_0$, while the horizon coordinates $r_+$ and $r_-$ (and hence the angular momentum) are allowed to vary. As we now proceed to show, smoothness at the horizon imposes a constraint on the allowed values of $r_+$ and $r_-$. Following \cite{Banados:1992wn, Mann:1996bi}, we consider the metric \eqref{eq:rotatingBTZMetric} in the near-horizon region, for $r\to r_+$. It is useful to introduce a new radial coordinate $\rho$ such that in the near-horizon region, we have
\begin{equation}
\begin{split}
    &\Delta(r) \approx \Delta'(r_+)(r-r_+) \equiv 2\kappa(r-r_+) = \kappa^2 \rho^2\,, \\
    &(r-r_+) = \frac{\kappa}{2}\rho^2\,, \quad \kappa = r_+(r_+^2-r_-^2) \,, \quad dr = \kappa\rho d\rho \,.
\end{split}
\end{equation}
Let
\begin{equation}
    \psi = \phi' + \ii\left(\frac{r_-}{r_+} - \Omega_0\right)t_E \,.
\end{equation}
Then the metric in the NHR up to terms $\O(\rho^2)$ is given by
\begin{equation} \label{eq:BTZ partial decomp}
    ds^2 = ds_\Sigma^2 + r_+^2\left(d\rho^2 + \frac{\kappa^2 \rho^2}{r_+^4} dt_E^2\right) \,,\quad ds_\Sigma^2 = r_+^2d\psi^2
\end{equation}
where $ds_\Sigma^2$ is evaluated on the horizon surface $\Sigma$. The metric \eqref{eq:BTZ partial decomp} may be rewritten as
\begin{equation}
    ds^2 = ds_\Sigma^2 + r_+^2 ds_{C_2}^2\,,\quad ds_{C_2}^2 = d\rho^2 + \frac{\kappa^2 \rho^2}{r_+^4} dt_E^2
\end{equation}
where just as in \cite{Mann:1996bi}, $ds_{C_2}^2$ is the metric of a 2D disk $C_2$ attached to the horizon at $\psi$. In order to avoid a conical singularity at the horizon, we impose the periodicity
\begin{equation} \label{eq:BTZ psi periodicity}
    (t_E, \psi) \sim \left(t_E + \frac{2\pi r_+}{r_+^2-r_-^2}, \psi\right) \,,
\end{equation}
from which we define $\beta\equiv \frac{2\pi r_+}{r_+^2-r_-^2}$. Plugging back in our original identification, \eqref{eq:BTZ psi periodicity} takes the form
\begin{equation}
    \left(t_E, \phi' + \ii\left(\frac{r_-}{r_+} - \Omega_0\right)t_E\right) \sim \left(t_E+\beta, \phi' - \ii\left(\frac{r_-}{r_+} - \Omega_0\right)\beta + \ii\left(\frac{r_-}{r_+} - \Omega_0\right)(t_E+\beta)\right)
\end{equation}
This requires a simultaneous shift in $t_E$ and $\phi'$ in order for $\psi$ to remain fixed as $t_E$ goes around the thermal circle,
\begin{equation}
    (t_E,\phi') \sim \left(t_E + \beta, \phi' - \ii\left(\frac{r_-}{r_+} - \Omega_0\right)\beta\right) \,.
\end{equation}
We now proceed to check that this solution satisfies the boundary conditions on the fermions set by Eq.~\eqref{eq:fermionBC3d}. Since the fermions in this geometry are $4\pi$-periodic, any multiple of $4\pi$ labeled by $n\in\Z$ corresponds to a distinct solution for the parameters $r_+$ and $r_-$. In particular, fermions in this geometry satisfy the periodicity
\begin{equation}
    \psi(t_E, r, \phi') = -\psi\left(t_E+\beta, r, \phi' - \ii\left(\frac{r_-}{r_+} - \Omega_0\right)\beta\right) = -\psi\left(t_E+\beta, r, \phi' + 4\pi n\right)\,.
\end{equation}
We now see that Eq.~\eqref{eq:fermionBC3d} only holds if $r_+\equiv(r_+)_n$ and $r_-\equiv(r_-)_n$ satisfy the constraints\footnote{If we instead work in a theory without any fermionic degrees of freedom (see \footref{bosonic-vs-fermionic}), then \eqref{eq:3d theta periodicity} is replaced by
\begin{align}
     \frac{(r_-)_n}{(r_+)_n}= \Omega_0 + \frac{2\pi\ii n}{\beta}, \quad\text{and}\quad \beta = 2\pi \frac{(r_+)_n}{(r_+)_n^2-(r_-)_n^2}\,.
\end{align}  }

\begin{equation} \label{eq:3d theta periodicity}
    \frac{(r_-)_n}{(r_+)_n}= \Omega_0 + \frac{4\pi\ii n}{\beta}, \quad\text{and}\quad \beta = 2\pi \frac{(r_+)_n}{(r_+)_n^2-(r_-)_n^2}\,.
\end{equation}
Solving for $(r_\pm)_n$ as a function of boundary conditions $\beta$ and $\Omega_0$, we find that each $n$ results in a different BTZ solution, where the inner and outer horizons in \eqref{eq:rotatingBTZMetric} are given by
\begin{equation} \label{eq:rp rm}
    (r_+)_n = \frac{2\pi\beta}{\beta^2+(4\pi n-\ii\beta\Omega_0)^2}\,,\quad (r_-)_n = \frac{2\pi(4\pi\ii n+\beta\Omega_0)}{\beta^2+(4\pi n-\ii\beta\Omega_0)^2} \,.
\end{equation}
To summarize, the family of saddles is given by
\begin{equation} \label{eq:3dfamily}
    ds^2 = \frac{\Delta(r)}{r^2}\, dt_E^2 + \frac{r^2 dr^2}{\Delta(r)} + r^2 \left[d\phi' + \ii\left(\frac{(r_+)_n (r_-)_n}{r^2} - \Omega_0\right) dt_E \right]^2 \,,
\end{equation}
for $\Delta(r) = (r^2 - (r_+)_n^2)(r^2 - (r_-)_n^2)$ and where $(r_+)_n$ and $(r_-)_n$ are parametrized by \eqref{eq:rp rm}. Each choice of $(r_\pm)_n$ leads to a different angular momentum, mass, and entropy,
\begin{equation} \label{eq:JM_n}
    (J_{z,\text{cl}})_n = \frac{1}{4G_N}(r_+)_n(r_-)_n\,,\quad M_n = \frac{(r_+)_n^2+(r_-)_n^2}{8G_N}\,,\quad S_n = \frac{2\pi(r_+)_n}{4G_N} \,,
\end{equation}
which can be used to determine the on-shell action. Just as in the Maxwell case in \eqref{eq:AphiFixed}, the solutions \eqref{eq:3dfamily} are related to each other by a combination of a large gauge transformation with a shift in the parameters.

\subsubsection*{Quantization of angular momentum for BTZ}

In the fixed angular velocity ensemble, the quantization of angular momentum was previously observed in \cite{Maloney:2007ud}, though we rewrite it here as an illustration of the more general mechanism presented in \eqref{eq: general Dirichlet quantization}.

The partition function for the family of solutions \eqref{eq:3dfamily} is
\begin{align} \label{eq:BTZ one loop}
    Z_{\rm 3D}(\beta,&\Omega_0) \supset \sum_{n\in\Z} e^{-I_{\text{cl}} \left(\beta,\Omega_0+\frac{4\pi\ii}{\beta}n\right)} Z_{\text{1-loop}}\left(\beta,\Omega_0+\frac{4\pi\ii}{\beta}n\right) \,,\\
    &I_{\text{cl}}\left(\beta,\Omega_n\right) = \beta M_n-\beta\Omega_n(J_{z,\text{cl}})_n-S_n = -\frac{\ii\pi}{8G_N} \left(\frac{1}{\tau-2n} - \frac{1}{\bar\tau-2n}\right)\,,\nonumber \\
    &Z_{\text{1-loop}}\left(\beta,\Omega_n\right) = \prod_{m=2}^{\infty} \left(1-e^{-\ii\frac{2\pi}{\tau-2 n}m}\right)^{-1} \left(1-e^{\ii\frac{2\pi}{\bar\tau-2 n}m}\right)^{-1} \,, \nonumber
\end{align}
where $\tau = \frac{\ii}{2\pi}(\beta\Omega_0+\beta)$, $\bar\tau = \frac{\ii}{2\pi}(\beta\Omega_0-\beta)$, and where $M_n$, $(J_{z,\text{cl}})_n$, and $S_n$ are functions of $\beta$, $\Omega_0$, and $n$ given by \eqref{eq:JM_n}.\footnote{Note that for the family of saddles parametrized by $n$, the parameters $M$, $J_{z,\text{cl}}$, and $S$ cannot be identified as the thermodynamic mass, angular momentum, and entropy unless one saddle dominates. This is consistent with the boundary conditions that we impose since $M$, $J_{z,\text{cl}}$, and $S$ are allowed to vary while we keep $\beta$ and $\Omega_0$ fixed.} The one-loop fluctuations were computed in \cite{Maloney:2007ud, Giombi:2008vd}. Applying Poisson resummation to \eqref{eq:BTZ one loop}, we directly see that this family of contributions to the partition function only has support for half-integer angular momentum,
\begin{equation} \label{eq:BTZ poisson}
    Z_{\rm 3D}(\beta,\Omega_0) \supset \sum_{J_z\in\frac12\Z} e^{\beta\Omega_0J_z} Z_{\rm BTZ}(\beta,J_z)\,.
\end{equation}

By rewriting the partition function \eqref{eq:BTZ one loop} as
\begin{equation}
    Z_{\rm 3D}(\beta,\Omega_0) \supset \sum_{\gamma\in \{ST^{2n}\}} Z_{\text{TAdS}}(\gamma\tau, \gamma\bar\tau) \,,
\end{equation}
we see that we have only included a subset of the classical saddles compared to \eqref{eq:Gtheta}. More generally, the family of shifted saddles in \eqref{eq:3dfamily} corresponds to the right coset by $\Z$, which identifies $\gamma\sim \gamma T^{2n}$ for $n\in\Z$:
\begin{equation}
    Z_{\rm 3D}(\beta,\Omega_0) = Z_{\text{TAdS}}(\tau, \bar\tau) + \sum _{\Z\backslash\Gamma_\theta/\Z} ~\sum_{n\in\Z} Z_{\text{TAdS}}(\gamma T^{2n}\tau, \gamma T^{2n}\bar\tau) \,.
\end{equation}
For each $\gamma\neq I$, Poisson resummation results in the corresponding family of fixed (half-integer) angular momentum saddles.

In the next subsection, we will show the family of saddles that contribute to $Z_{\rm BTZ}(\beta,J_z)$.

\subsection*{3D saddles at fixed angular momentum}
We now consider solutions to the fixed $J_z$ Euclidean path integral in AdS$_3$ with action:
\begin{equation}
    I_{\text{EH}}(\beta,J_z) = I_{\text{EH}}(\beta,\Omega_0) + \int_{\p\M_3}\sqrt{\sigma} V^\phi j_\phi + I_{\rm ct}\,,
\end{equation}
where we add the boundary term from \eqref{eq: Brown-York bdy term} to the action \eqref{eq:BTZ Action}. In addition to fixing $j_\phi$, we impose the boundary topology $T^2$ with asymptotic metric
\begin{equation}
    ds^2 = r^2dt_E^2 + \frac{dr^2}{r^2} + r^2(d\phi' - \ii\Omega dt_E)^2 + \ldots\,,\quad r\to\infty \,,
\end{equation}
where $\Omega$ is a free parameter determined by $V^\phi$ as in \eqref{eq:Omega_n from V^phi}.

The family of saddles at fixed angular momentum takes a similar form to Eq.~\eqref{eq:rotatingBTZMetric}, where now $r_+$ and $r_-$ are fixed while the order $r^2$ coefficient of $d\phi dt_E$ is free. Explicitly, the solutions take the form
\begin{equation}
    ds^2 = \frac{\Delta(r)}{r^2}\, dt_E^2 + \frac{r^2 dr^2}{\Delta(r)} + r^2 \left[d\phi' + \ii\left(\frac{r_+ r_-}{r^2} - \Omega\right) dt_E \right]^2\,, \quad \Delta(r) = (r^2 - r_+^2)(r^2 - r_-^2) \,,
\end{equation}
where the analogue of the smoothness condition \eqref{eq:3d theta periodicity} now results in $\Omega\equiv\Omega_n$ for
\begin{equation} \label{eq:3D neumann velocities}
    \Omega_n = \frac{r_-}{r_+} + \frac{4\pi\ii n}{\beta} \,,\quad n\in\Z \,.
\end{equation}
These are the saddles we summed over in the saddlepoint approximation of \eqref{eq:brownyork fixedJ}, with $\Omega_0 = \frac{r_-}{r_+}$. Because $r_\pm$ are fixed, the different $n$-images describe the same bulk BTZ geometry, with the same $M$, $J_z$, and $S$; the only $n$-dependence is the boundary phase $e^{4\pi\ii nJ_z}$, whose regulated overall normalization is removed by $I_{\rm ct}$, analogously to \eqref{eq:mc Z}.

The fixed $J_z$ partition function can be found from the Fourier transform of \eqref{eq:BTZ one loop}:
\begin{equation}
    Z_{\rm BTZ}(\beta,J_z) = \frac{1}{4\pi} \int_{-\infty}^{\infty} d\theta e^{-\ii\theta J_z} e^{-I_{\text{cl}}\left(\beta,\Omega\right)} Z_{\text{1-loop}}\left(\beta,\Omega\right) \approx e^{S - \beta M} Z_{\text{1-loop}}(\beta,J_z) (1+\ldots) \sum_{n\in\Z} e^{4\pi\ii n J_z} \,,
\end{equation}
with saddlepoints at $\theta=-\ii\beta\Omega_n$ and just as in \eqref{eq: general Neumann quantization}, this simplifies to
\begin{equation}
    Z_{\rm BTZ}(\beta,J_z) = \left\{ \begin{matrix} e^{S - \beta M} Z_{\text{1-loop}}(\beta,J_z) (1+\ldots) & J_z\in\frac12\Z \\ 0 & \text{else} \end{matrix} \right. \,.
\end{equation}
The on-shell fixed $\beta,J_z$ action simplifies to
\begin{equation}
    -I_{\rm BTZ}(\beta,J_z) = S - \beta M = \frac{2\pi r_+}{4G_N} - \beta\frac{r_+^2+r_-^2}{8G_N}
\end{equation}
with the values of $r_+,r_-$ determined from $\beta$ and $J_z$ via \cite{Banados:1992wn}
\begin{equation}
    J_z = \frac{1}{4G_N}r_+r_-\,,\quad \beta = 2\pi\frac{r_+}{r_+^2-r_-^2} \,.
\end{equation}

\section{4D gravity} \label{sec: 4d}

We next apply the lessons of the pedagogical discussion above to $4D$ black holes. In the main text, we will focus on asymptotically flat saddles, while in Appendix \ref{sec:AdS4}, we similarly construct an infinite set of AdS saddles. 
For a fixed choice of $\Omega_0$ or $J_z$, something special happens in 3D: the saddles were related by large boundary diffeomorphisms. These diffeomorphisms change the modular parameter $\tau$ of the boundary, and they act on $\tau$ as $T^{2n}$-transforms, which are a subset of the larger $\Gamma_\theta\subset SL(2,\Z)$ mapping class group of the dual 2D CFT. However, as we now illustrate in 4D, the existence of a family of fixed $\Omega_0$ or $J_z$ saddles is unrelated to the boundary mapping class group.\footnote{Something analogous happens in the Maxwell example that we discussed in Section \ref{section: maxwell}. Our vector potential transformed under the $U(1)$ gauge group, and naively, we might have guessed that the family of saddles given by \eqref{eq:OnShellAction} should correspond to the first homotopy group $\pi_1(U(1)) \cong\Z$. However, if instead we had considered an $SU(2)$ gauge field, we would have found the same family of saddles \eqref{eq:OnShellAction}, even though $\pi_1(SU(2))$ is trivial. Hence, the argument for the path integral quantization of charge is independent of the topology of the setup.} In this section, we find a family of 4D saddles analogous to the $T$-transforms in 3D that lead to the quantization of angular momentum.

\subsubsection*{``$T$-transformed'' Kerr-Newman saddles at fixed angular velocity}
We consider solutions to the Euclidean path integral in asymptotically flat space with Einstein-Maxwell action
\begin{equation} \label{eq:EM4dflat}
\begin{split}
    I_{\rm EM}(\beta,\Omega_0,Q) = -\frac{1}{16\pi G_N}&\left[\int_{\M_4} d^4x\sqrt{g}R - 2\int_{\p\M_4}\sqrt{h}K\right] \\
    &- \frac{1}{4e^2}\int_{\M_4}d^4x\sqrt{g}F_{\mu\nu}F^{\mu\nu} - \frac{1}{e^2}\int_{\p\M_4}d^3x\sqrt{h}A_\mu F^{\mu\nu}n_\nu\,, 
\end{split}
\end{equation}
where $F=dA$, $A$ is pure imaginary, and $\Omega_0$ is in Euclidean signature.\footnote{If we instead worked at fixed electric potential, the analogous grand-canonical boundary conditions were discussed in \cite{Braden:1990hw, Hawking:1995ap}. Here we work with boundary conditions in which the field strength is fixed, therefore fixing the charge to be integer quantized.} We consider the boundary topology $S^1\times S^2$ and asymptotic metric
\begin{equation}
    ds^2 = d\tau^2 + dr^2 + r^2d\theta^2 + r^2\sin^2\theta (d\phi' - \ii\Omega_0 d\tau)^2 + \ldots\,,
\end{equation}
where we fix the angular velocity $\Omega_0$ at the boundary. The corotating coordinate $\phi'$ is related to the usual Boyer-Lindquist coordinate $\phi$ via the diffeomorphism
\begin{equation} \label{eq:corotating}
    \phi' = \phi + \ii\Omega_0\tau\,.
\end{equation}
Just as in 3D, we impose the following boundary conditions on fermionic fields propagating on this background,\footref{bosonic-vs-fermionic}
\begin{equation}\label{eq:4dfermionBC}
    \psi(\tau, r,\theta, \phi') = -\psi(\tau+\beta, r,\theta, \phi') = \psi(\tau, r,\theta, \phi'+4\pi) \,.
\end{equation}
Consider the family of classical saddles at fixed angular velocity $\Omega_0$, given by
\begin{equation} \label{eq:4dkerrnewman}
    ds^2 = \frac{\rho^2\Delta}{\Sigma}d\tau^2 + \frac{\rho^2}{\Delta}dr^2 + \rho^2 d\theta^2 + \sin^2\theta\frac{\Sigma}{\rho^2}\left[d\phi' + \left(\frac{\hat a}{\Sigma}(2mr-Q^2) - \ii\Omega_0\right) d\tau\right]^2
\end{equation}
where the mass is parametrized as $M=\frac{m}{G_N}$ and $\hat a$ is the angular momentum parameter in Euclidean signature, which is left unfixed.\footnote{This Euclidean angular momentum is related to the Lorentzian one $a$ via $\hat a=\ii a$. Notice that analytic continuation of Euclidean $a,\Omega_0,\tau$ leads to a real Lorentzian metric as well.} The other functions appearing in the metric \eqref{eq:4dkerrnewman} are given by
\begin{equation}
    \rho^2 = r^2-\hat a^2\cos^2\theta\,,\quad \Delta = r^2-\hat a^2-2mr+Q^2\,,\quad \Sigma = (r^2-\hat a^2)^2 + \hat a^2\Delta\sin^2\theta\,,
\end{equation}
and near the asymptotic boundary, the cross term takes the form
\begin{equation}
    \lim_{r\to\infty} g_{\tau\phi'} = -\ii\Omega_0(r^2 - \hat a^2) \sin^2\theta\,.
\end{equation}
Note that the coefficient of $r^2$ determines the angular velocity $\Omega_0$ while the terms subleading in $r$ determine the angular momentum $J=Ma$. The metric has an outer and inner event horizon at radii
\begin{equation} \label{eq:4d event horizons}
    r_\pm = m\pm\sqrt{m^2-Q^2+\hat a^2} \,,
\end{equation}
whose dependence on boundary conditions we will determine shortly.

We follow the procedure outlined in \cite{Mann:1996bi} to check the smoothness of these solutions. First, we introduce a new set of coordinates $(x,\chi,\theta,\psi)$ given by 
\begin{equation} \label{eq:4d rindler coord}
    x^2 = \frac{4}{\gamma}(r-r_+)\,,\quad \chi = \tau-\hat a (\phi'-\ii\Omega_0\tau) \sin^2\theta\,,\quad \psi = \phi' + \left(\frac{\hat a}{r_+^2-\hat a^2} - \ii\Omega_0\right) \tau \,,
\end{equation}
where $\gamma = \Delta'(r_+) = 2\sqrt{m^2-Q^2+\hat a^2}$.
Near the horizon, we write the metric \eqref{eq:4dkerrnewman} up to terms of $\O(x^2)$:
\begin{equation} \label{eq:4d near horizon}
    ds^2 = ds_\Sigma^2 + \rho_+^2 ds_{C_2}^2\,,\quad ds_{C_2}^2 = dx^2 + \frac{\gamma^2x^2}{4\rho_+^4}d\chi^2\,,\quad ds_\Sigma^2 = \rho_+^2d\theta^2 + \frac{(r_+^2-\hat a^2)^2}{\rho_+^2}\sin^2\theta d\psi^2\,,
\end{equation}
where $ds_\Sigma^2$ is the metric on the horizon surface $\Sigma$ (at $r=r_+$), $ds_{C_2}^2$ is the metric of a disk $C_2$ attached to $\Sigma$ at a point $(\theta,\psi)$, and $\rho_+$ is the coordinate $\rho$ evaluated at $r=r_+$. The coordinate $\chi$ forms a well-defined angle on $C_2$, while $\psi$ forms a well-defined angle on $\Sigma$. At fixed $(\theta,\psi)$, for the metric $ds_{C_2}^2$ to be smooth at the horizon, i.e. near $x=0$, we impose the periodicity
\begin{equation}
    \chi \sim \chi + \frac{4\pi}{\gamma}\rho_+^2 \,. \label{eq: KN chi}
\end{equation}
In terms of the $(\tau,\phi')$ coordinates, this translates into
\begin{equation}
\begin{split}
    \tau - \hat a(\phi'-\ii\Omega_0\tau)\sin^2\theta &\sim \tau - \hat a(\phi'-\ii\Omega_0\tau)\sin^2\theta + \frac{4\pi}{\gamma}\left[r_+^2-\hat a^2(1-\sin^2\theta)\right] \\
    &\sim \tau + \beta - \hat a\sin^2\theta\left(\phi' - \ii\Omega_0(\tau+\beta) + \ii\Omega_0\beta - \frac{4\pi \hat a}{\gamma}\right) \label{eq: KN Chi Algebra}
\end{split}
\end{equation}
where we defined $\beta\equiv \frac{4\pi}{\gamma}(r_+^2-\hat a^2)$. Then a smooth geometry at the horizon requires
\begin{equation}
    (\tau,\phi') \sim \left(\tau+\beta,\phi' + \ii\Omega_0\beta - \frac{\hat a}{r_+^2-\hat a^2}\beta\right) \,.
\end{equation}

Since fermions in this geometry are again $4\pi$-periodic (and $2\pi$ anti-periodic) in $\phi'$, there exists a family of distinct solutions for $r_+$ and $a$ labeled by $n\in\Z$. Recall the periodicity obeyed by fermions in this geometry is given by
\begin{equation}
    \psi(\tau, r,\theta, \phi') = -\psi\left(\tau+\beta, r,\theta, \phi' + \ii\Omega_0\beta - \frac{\hat a}{r_+^2-\hat a^2}\beta\right) = -\psi\left(\tau+\beta, r,\theta, \phi' + 4\pi n\right) \,.
\end{equation}
Eq.~\eqref{eq:4dfermionBC} is satisfied if and only if $\hat a\equiv \ii a_n$ and $r_+\equiv (r_+)_n$  are solutions to the following system
\begin{equation} \label{eq:4d omega periodicity}
    \frac{a_n}{(r_+)_n^2 + a_n^2} = \Omega_0 + \frac{4\pi\ii n}{\beta} \,,\quad \beta = \frac{4\pi (r_+)_n[(r_+)_n^2 + a_n^2]}{(r_+)_n^2 - a_n^2 - Q^2} \,,
\end{equation}
from which we see that the angular momentum parameter $a_n$ and the outer horizon $(r_+)_n$ both depend on $n$. The angular momentum, mass, and entropy of the Kerr-Newman black hole,
\begin{equation} \label{eq:JMn_4d}
    (J_{z,\text{cl}})_n = M_na_n\,,\quad M_n = \frac{(r_+)_n^2+Q^2+a_n^2}{2G_N (r_+)_n}\,,\quad S_n = \frac{\pi[(r_+)_n^2+a_n^2]}{G_N}\,,
\end{equation}
are also distinct for each $n$. The action is given by $-I_n = S_n +\beta \Omega_0 (J_{z,\text{cl}})_n - \beta M_n$ and is also explicitly $n$-dependent. Hence, for a fixed $\Omega_0$ boundary condition, the family of solutions is parametrized by $a_n$ and $(r_+)_n$, where each choice of $(a_n,(r_+)_n)$ leads to different $(J_{z,\text{cl}})_n,M_n,S_n$. Just as in 3D, we can plug these saddles into the path integral \eqref{eq: general Dirichlet quantization} and observe the quantization $J_z\in\frac12\Z$.\footnote{If we instead work in a theory without any fermionic degrees of freedom (see \footref{bosonic-vs-fermionic}), then \eqref{eq:4d omega periodicity} is replaced by
\begin{align}
     \frac{a_n}{(r_+)_n^2 + a_n^2} = \Omega_0 + \frac{2\pi\ii n}{\beta} \,,\quad \beta = \frac{4\pi (r_+)_n[(r_+)_n^2 + a_n^2]}{(r_+)_n^2 - a_n^2 - Q^2}\,.
\end{align}  
In this case, the partition function can be expressed as a sum over $SO(3)$ irreps rather than $SU(2)$. Additionally note that both in the cases with and without fermionic degrees of freedom, the classical angular momentum associated with each saddle, $(J_{z,\text{cl}})_n$ is not quantized even though $J_z$ is. 
}  Whereas the 3D gravity $T$-transforms were part of the torus mapping class group $SL(2,\Z)$, the mapping class group of $S^1\times S^2$ is $\Z_2$ \cite{bams/1183524395}, which does not contain the infinite family of solutions that we explicitly wrote down above. As a consequence, the $T$-transformed 3D saddles were related by a large boundary diffeomorphism, while the ``$T$-transformed'' 4D saddles are not.

The construction above exhibits an infinite family of smooth Euclidean solutions with the same Dirichlet boundary data; however, it does not by itself determine which of these saddles lie on the relevant integration contour. In Appendix \ref{app: convergence}, we check the weaker question of whether the sum over $n$ is convergent when the full family is included; the answer is branch- and parameter-dependent. In particular, the charged asymptotically flat Kerr-Newman solution has convergent large $n$ asymptotics, while the uncharged and AdS cases require additional input, such as one-loop effects or a contour prescription.

Note that in the extremal limit, the difference between the actions of many saddles can be very small. This is the case for Kerr-Newman black holes whose contribution can, for instance, be isolated by taking $\Omega_0$ to be purely imaginary and $\beta \to \infty$. Because the saddles contributing to the fixed angular velocity ensemble have nearly degenerate actions in this limit, the full sum over saddles is required to correctly compute even basic observables in the ensemble, such as $\braket{J_z}$ or $\braket{J^2_z}$. We explicitly demonstrate this in Appendix \ref{app:variance}, where we also study how the summation over saddles affects the extremal limits of other black holes.

\subsubsection*{``$T$-transformed'' Kerr-Newman saddles at fixed angular momentum}
We now consider solutions to the fixed $J_z$ Euclidean path integral in asymptotically flat space with action:
\begin{equation}
    I_{\rm EM}(\beta,J_z,Q) = I_{\rm EM}(\beta,\Omega_0,Q) + \int_{\p\M_4}\sqrt{\sigma} V^aj_a + I_{\rm ct} \,,
\end{equation}
where we add the boundary term from \eqref{eq: Brown-York bdy term} to the action \eqref{eq:EM4dflat}. In addition to fixing $j_\phi$, we impose the boundary topology $S^1\times S^2$ with asymptotic metric
\begin{equation}
    ds^2 = d\tau^2 + dr^2 + r^2d\theta^2 + r^2\sin^2\theta (d\phi' - \ii\Omega d\tau)^2 + \ldots\,,
\end{equation}
where the free parameter $\Omega$ is determined by $V^\phi$ using \eqref{eq:Omega_n from V^phi}, and fermions satisfy
\begin{equation} \label{eq:4d neumann fermions}
    \psi(\tau, r,\theta, \phi') = -\psi(\tau+\beta, r,\theta, \phi') = -\psi(\tau+\beta, r,\theta, \phi' + 4\pi)\,.
\end{equation}
Just as in 3D, the family of solutions at fixed angular momentum takes a similar form to Eq.~\eqref{eq:4dkerrnewman}, where instead the angular momentum parameter $\hat a_0$ is fixed (which holds fixed the location of the event horizons $r_\pm$ in \eqref{eq:4d event horizons} while the order $r^2$ coefficient of $d\phi d\tau$ is free to fluctuate). The saddles have a metric that takes the form
\begin{equation}
    ds^2 = \frac{\rho^2\Delta}{\Sigma}d\tau^2 + \frac{\rho^2}{\Delta}dr^2 + \rho^2 d\theta^2 + \sin^2\theta\frac{\Sigma}{\rho^2}\left[d\phi' + \left(\frac{\hat a_0}{\Sigma}(2mr-Q^2) - \ii\Omega\right) d\tau\right]^2
\end{equation}
and smoothness at the black hole horizon combined with the boundary condition \eqref{eq:4d neumann fermions} imposes the following periodicity on fermions,
\begin{equation}
    \psi(\tau, r,\theta, \phi') = -\psi\left(\tau+\beta, r,\theta, \phi' + \ii\Omega\beta - \frac{4\pi \hat a_0}{\gamma}\right) = -\psi\left(\tau+\beta, r,\theta, \phi' + 4\pi n\right) \,.
\end{equation}
This implies $\Omega\equiv\Omega_n$ is given by
\begin{equation} \label{eq:4d Neumann velocities}
    \Omega_n = \frac{a_0}{r_+^2 + a_0^2} - \frac{4\pi\ii n}{\beta} \,,\quad n\in\Z \,,
\end{equation}
where $a_0=-\ii\hat a_0$. Now $\Omega_n$ depends on $n$, while $r_+$ and $M$ are fixed. It is immediate from \eqref{eq: general Neumann quantization} that $J_z$ is quantized for this ensemble of 4D Kerr-Newman saddles. 

\section*{Acknowledgements}
We thank Maciej Kolanowski, Don Marolf, Rob Myers, Joaquin Turiaci, and especially Jan Boruch and Guanda Lin for valuable discussions. LVI was supported in part by the Leinweber Institute for Theoretical Physics
at UC Berkeley, by the Department of Energy, Office of Science, Office of High Energy
Physics through the award DE-SC0025522, and by the Department of
Energy through QuantISED award DE-SC0019380. 

\appendix
\section{Einstein-Maxwell in asymptotically AdS$_4$} \label{sec:AdS4}
In this section, we generalize the 4D Kerr-Newman results to AdS$_4$: we briefly discuss the fixed angular velocity and momentum boundary conditions and show the resulting family of saddles.

\subsubsection*{AdS$_4$ saddles at fixed angular velocity}
We consider solutions to the Euclidean path integral in AdS$_4$ where the Einstein-Maxwell action now includes the cosmological constant,
\begin{align}
    I_{\rm EM}(\beta,\Omega_0,Q) = -\frac{1}{16\pi G_N}&\left[\int_{\M_4} d^4x\sqrt{g}\left(R - 2\Lambda\right)-2\int_{\p\M_4}\sqrt{h}K\right] \nonumber\\
    &- \frac{1}{4e^2}\int_{\M_4}d^4x\sqrt{g}F_{\mu\nu}F^{\mu\nu} - \frac{1}{e^2}\int_{\p\M_4}d^3x\sqrt{h}A_\mu F^{\mu\nu}n_\nu\,,
\end{align}
where $\Lambda = \frac{-3}{\ell^2}$ and $\ell$ is the AdS length. We impose the condition \eqref{eq:4dfermionBC} on the fermions as well as the boundary topology $S^1\times S^2$ with asymptotic metric
\begin{equation}
    ds^2 = \frac{\Delta_\theta}{\Xi} \frac{r^2}{\ell^2} d\tau^2 + \frac{\ell^2}{r^2}dr^2 + \frac{r^2}{\Delta_\theta} d\theta^2 + \frac{r^2}{\Xi}\sin^2\theta \left[d\phi' - \ii\Omega_0 d\tau\right]^2 + \ldots\,,
\end{equation}
where $\Delta_\theta$ and $\Xi$ are defined in \eqref{eq:ads4 coordinates} below. The coordinate $\phi'$ is corotating with respect to the non-rotating frame at infinity:
\begin{equation} \label{eq:AdS4 corotating}
    \phi' = \phi + \ii\Omega_0\tau - \ii \frac{a}{\ell^2}\tau = \phi + \ii\Omega_H\tau\,.
\end{equation}
AdS$_4$ is distinct from the cases we have considered in the main text so far in that the angular velocity at the horizon, $\Omega_H$, which defines the Killing horizon generator
\begin{equation}
    \xi = \p_\tau - \ii\Omega_H\p_\phi\,,
\end{equation}
differs from the parameter $\Omega_0$ conjugate to the thermodynamic angular momentum $J_z$ \cite{Caldarelli:1999xj, Gibbons:2004ai}. The angular velocity $\Omega_0$ is measured relative to the non-rotating frame at infinity and is hence defined by $\Omega_0 = \Omega_H-\Omega_\infty = \Omega_H+a/\ell^2$, where $\Omega_\infty$ is the angular velocity at asymptotic infinity. With respect to the coordinate $\phi'$ defined by \eqref{eq:AdS4 corotating}, fermionic fields in this background satisfy the boundary conditions
\begin{equation} \label{eq:AdS4 fermionBC}
    \psi(\tau, r,\theta, \phi') = -\psi(\tau+\beta, r,\theta, \phi') = \psi(\tau, r,\theta, \phi'+4\pi)\,.
\end{equation}

The family of classical saddles at fixed angular velocity $\Omega_0$ is given by
\begin{equation} \label{eq:ads4kerrnewman}
    ds^2 = \frac{\rho^2\Delta_r\Delta_\theta}{\Sigma} d\tau^2 + \frac{\rho^2}{\Delta_r}dr^2 + \frac{\rho^2}{\Delta_{\theta}}d\theta^2 + \sin^2\theta \frac{\Sigma}{\rho^2\Xi^2}\left[d\phi'+\left(\frac{\hat a\Xi}{\Sigma}[(r^2-\hat a^2)\Delta_\theta - \Delta_r] + \frac{\hat a}{\ell^2} - \ii\Omega_0\right)d\tau\right]^2
\end{equation}
where $\hat a=\ii a$, $M=\frac{m}{G_N\Xi^2}$, $Q=\frac{q}{\Xi}$, and 
\begin{equation} \label{eq:ads4 coordinates}
\begin{split}
    &\rho^2 = r^2-\hat a^2\cos^2\theta\,,\quad \Delta_r = (r^2-\hat a^2) \left(1+\frac{r^2}{\ell^2}\right)-2mr + q^2\,,\\
    &\Delta_\theta = 1+\frac{\hat a^2}{\ell^2}\cos^2{\theta},\quad \Xi = 1+\frac{\hat a^2}{\ell^2}\,,\quad \Sigma = (r^2-\hat a^2)^2\Delta_\theta + \hat a^2\Delta_r\sin^2\theta\,.
\end{split}
\end{equation}
In the limit $\ell\to\infty$ these reduce to the asymptotically flat family of Kerr-Newman saddles \eqref{eq:4dkerrnewman}.

We can carry out an analogous check of smoothness to the one in asymptotically flat Kerr-Newman, with the change of coordinates that takes us to the near-horizon region:
\begin{equation}
\begin{split}
    x^2 = \frac{4}{\gamma}(&r-r_+)\,,\quad \chi = \tau - \frac{\hat a}{\Xi}\left(\phi' - \ii\Omega_0\tau + \frac{\hat a}{\ell^2}\tau\right)\sin^2\theta\,, \\
    &\psi = \phi' - \ii\Omega_0\tau + \frac{\hat a}{\ell^2}\tau + \frac{\hat a\Xi}{r_+^2-\hat a^2} \tau \,,
\end{split}
\end{equation}
where $\gamma\equiv\Delta_r'(r_+)$ and $r_+$ is the radius of the outer event horizon. This results in the near-horizon metric
\begin{equation} \label{eq:ads near horizon}
    ds^2 = ds_\Sigma^2 + \rho_+^2 ds_{C_2}^2\,,\quad ds_{C_2}^2 = dx^2 + \frac{\gamma^2x^2}{4\rho_+^4}d\chi^2\,,\quad ds_\Sigma^2 = \frac{\rho_+^2}{\Delta_\theta} d\theta^2 + \frac{\Delta_\theta}{\Xi^2} \frac{(r_+^2-\hat a^2)^2}{\rho_+^2}\sin^2\theta d\psi^2\,,
\end{equation}
where we use smoothness in the $(x,\chi)$ coordinates to solve for $r_+$ and $\hat a$. The periodicity imposed on fermions is given by
\begin{equation}
    \psi(\tau, r,\theta, \phi') = -\psi\left(\tau+\beta, r,\theta, \phi' + \ii\Omega_0\beta - \frac{\hat a}{\ell^2}\beta - \frac{\hat a \Xi}{r_+^2 - \hat a^2}\beta\right) = -\psi\left(\tau+\beta, r,\theta, \phi' + 4\pi n\right) \,,
\end{equation}
resulting in the system of equations
\begin{equation} \label{eq:ads4 omega periodicity}
    \frac{a_n[(r_+)_n^2 + \ell^2]}{\ell^2[(r_+)_n^2 + a_n^2]} = \Omega_0 + \frac{4\pi\ii n}{\beta} \,,\quad \beta = \frac{4\pi[(r_+)_n^2 + a_n^2]}{(r_+)_n\left(1 + \frac{1}{\ell^2} [a_n^2+3(r_+)_n^2] - \frac{a_n^2+Q^2(1-a_n^2/\ell^2)^2}{(r_+)_n^2}\right)} \,.
\end{equation}
where $\hat a\equiv \ii a_n$ and $r_+\equiv (r_+)_n$ now depend on the value of $n\in\Z$. For $n=0$, these periodicity conditions match with those found in \cite{Caldarelli:1999xj}. The angular momentum, mass, and entropy of each AdS$_4$ Kerr-Newman black hole saddle is given by
\begin{equation} \label{eq:JMn_AdS4}
    (J_{z,\text{cl}})_n = M_na_n\,,\quad M_n = \frac{[(r_+)_n^2+a_n^2]\left(1+\frac{(r_+)_n^2}{\ell^2}\right) + \Xi_n^2Q^2}{2G_N (r_+)_n\Xi_n^2} \,,\quad S_n = \frac{\pi[(r_+)_n^2+a_n^2]}{G_N\Xi_n}
\end{equation}
as a function of $\beta$, $\Omega_0$, $Q$, and $n$, where $\Xi_n=1-\frac{a_n^2}{\ell^2}$.

\subsubsection*{AdS$_4$ saddles at fixed angular momentum} 

We now consider solutions to the fixed $J_z$ Euclidean path integral in asymptotically AdS$_4$. The asymptotic metric is given by
\begin{equation}
    ds^2 = \frac{\Delta_\theta}{\Xi} \frac{r^2}{\ell^2} d\tau^2 + \frac{\ell^2}{r^2}dr^2 + \frac{r^2}{\Delta_\theta} d\theta^2 + \frac{r^2}{\Xi}\sin^2\theta \left[d\phi' - \ii\Omega d\tau\right]^2 + \ldots\,,
\end{equation}
where $\Omega$ is a free parameter. The family of saddles takes the form
\begin{equation}
    ds^2 = \frac{\rho^2\Delta_r\Delta_\theta}{\Sigma} d\tau^2 + \frac{\rho^2}{\Delta_r}dr^2 + \frac{\rho^2}{\Delta_{\theta}}d\theta^2 + \sin^2\theta \frac{\Sigma}{\rho^2\Xi^2}\left[d\phi' + \left(\frac{\hat a_0\Xi}{\Sigma}[(r^2-\hat a_0^2)\Delta_\theta - \Delta_r] + \frac{\hat a_0}{\ell^2} - \ii\Omega\right)d\tau\right]^2
\end{equation}
for $\Omega\equiv\Omega_n$ given by
\begin{equation} \label{eq:ads4 Neumann velocities}
    \Omega_n = \frac{a_0(r_+^2+\ell^2)}{\ell^2(r_+^2 + a_0^2)} - \frac{4\pi\ii n}{\beta} \,,\quad n\in\Z \,.
\end{equation}
To find the quantization of the angular momentum $J_z$, we can now repeat the procedure in the main text.

\section{Convergence of the sum over saddles at fixed $\Omega$} \label{app: convergence}
In this appendix, we revisit the question about the convergence of the sum over saddles for fixed angular velocity in the cases of BTZ, asymptotically flat Kerr-Newman, and AdS$_4$ Kerr-Newman. We do not claim all saddles will end up contributing to the path integral (which would require establishing a correct contour prescription); rather we check whether the path integral converges if we do include the full family of saddles. In this appendix we let $G_N=1$.

\subsection{3D gravity partition function}

We would like to check the convergence of \eqref{eq:BTZ one loop}, repeated here for convenience:
\begin{equation}
\begin{split}
    Z_{\rm 3D}(\beta,\Omega_0) &\supset \sum_{n\in\Z} e^{-I_{\text{cl}}\left(\beta,\Omega_0+\frac{4\pi\ii}{\beta}n\right)} Z_{\text{1-loop}}\left(\beta,\Omega_0+\frac{4\pi\ii}{\beta}n\right) \\
    I_{\text{cl}}(\beta,\Omega_0) &= -\frac{\ii\pi}{8} \left(\frac{1}{\tau} - \frac{1}{\bar\tau}\right)\,,\quad Z_{\text{1-loop}}(\beta,\Omega_0) = \prod_{m=2}^{\infty} \left(1-e^{-\ii\frac{2\pi}{\tau}m}\right)^{-1} \left(1-e^{\ii\frac{2\pi}{\bar\tau} m}\right)^{-1} \,,
\end{split}
\end{equation}
where $\tau = \frac{\ii}{2\pi}(\beta\Omega_0+\beta)$ and $\bar\tau = \frac{\ii}{2\pi}(\beta\Omega_0-\beta)$. For large $n$, the on-shell action behaves as
\begin{equation}
    I_{\text{cl}}\left(\beta,\Omega_0+\frac{4\pi\ii}{\beta}n\right) = -\frac{\beta}{32} \frac{1}{n^2} + \O\left(\frac{1}{n^3}\right)
\end{equation}
so that $e^{-I_{\text{cl}}}\to1$ for large $n$. Without the one-loop determinant, the sum over saddles diverges. Now, accounting for the one-loop determinant, we have
\begin{equation}
    Z_{\text{1-loop}}\left(\beta,\Omega_0+\frac{4\pi\ii}{\beta}n\right) = \frac{e^{\beta/12}\pi^2}{2}\frac{1}{n^3} + \O\left(\frac{1}{n^4}\right)\,,
\end{equation}
which leads to a convergent sum.

\subsection{4D Kerr-Newman partition function}
We would like to check the convergence of
\begin{equation}
    Z_{\rm KN}(\beta,\Omega_0,Q) = \sum_{n\in\Z} e^{-I_{\rm KN}\left(\beta,\Omega_0+\frac{4\pi\ii}{\beta}n,Q\right)} Z_{\rm KN}^{(1)}\left(\beta,\Omega_0+\frac{4\pi\ii}{\beta}n,Q\right) \,,
\end{equation}
where the classical action of each saddle is given by Eq.~(2.11) of \cite{Iliesiu:2021are},
\begin{equation} \label{eq:app action}
    I_{\rm KN}\left(\beta,\Omega_0+\frac{4\pi\ii}{\beta}n,Q\right) = \frac{\beta}{2}\left(\frac{r_n^2+Q^2+a_n^2}{2r_n} + \frac{Q^2r_n}{r_n^2+a_n^2}\right)\,,
\end{equation}
for outer horizon $r_n = m_n+\sqrt{m_n^2-Q^2-a_n^2}$.

\textbf{Kerr black hole ($Q=0$)}: We first study the case of a non-charged, rotating black hole in asymptotically flat spacetime. The angular momentum parameter $a_n(\beta,\Omega_0)$ and outer horizon $r_n(\beta,\Omega_0)$ are solutions to the system of constraints given by \eqref{eq:4d omega periodicity} with $Q=0$ and fixed $\beta$ and $\Omega_0$ boundary conditions:
\begin{equation}
    \frac{a_n}{r_n^2+a_n^2} = \Omega_0 + \frac{4\pi\ii n}{\beta}\,,\quad \beta = \frac{4\pi r_n(r_n^2+a_n^2)}{r_n^2-a_n^2}\,,\quad n\in\Z \,.
\end{equation}
For any $n$, there exist two solutions to this system, parametrized by $\sigma=\pm1$,
\begin{align}
    &r_n = \sigma\frac{\beta}{2\Delta_n}\,,\quad a_n = \frac{1}{2\left(\Omega_0 + \frac{4\pi\ii n}{\beta}\right)} \left(1 - \sigma\frac{2\pi}{\Delta_n}\right) \label{eq:app_Q0_phys}
\end{align}
where 
$$\Delta_n = \sqrt{4\pi^2+\beta^2\left(\Omega_0 + \frac{4\pi\ii n}{\beta}\right)^2} \,.$$
We will refer to the solution \eqref{eq:app_Q0_phys}, with $\sigma=1$, as the physical branch. The action and mass for each $n$ are given by
\begin{equation*}
    I_{\text{Kerr}} \left(\beta,\Omega_0+\frac{4\pi\ii}{\beta}n\right)  = \frac{\beta m_n}{2} \,,\quad m_n = \frac{r_n^2+a_n^2}{2r_n}\,.
\end{equation*}
For large $n$, the action of the physical branch goes as
\begin{equation}
    I_{\text{Kerr}} \left(\beta,\Omega_0+\frac{4\pi\ii}{\beta}n\right) = -\frac{\ii\beta^2}{16\pi}\frac1n + \frac{2\pi\beta^2+\beta^3\Omega_0}{64\pi^2} \frac{1}{n^2} + \O\left(\frac{1}{n^3}\right) \,.
\end{equation}
Just as for BTZ, the sum over classical saddles diverges for $Q=0$, so we cannot conclude whether the sum converges or diverges without computing the one-loop determinant. 

\begin{figure}[h]
    \vspace{-.5cm}
    \includegraphics[width=\textwidth]{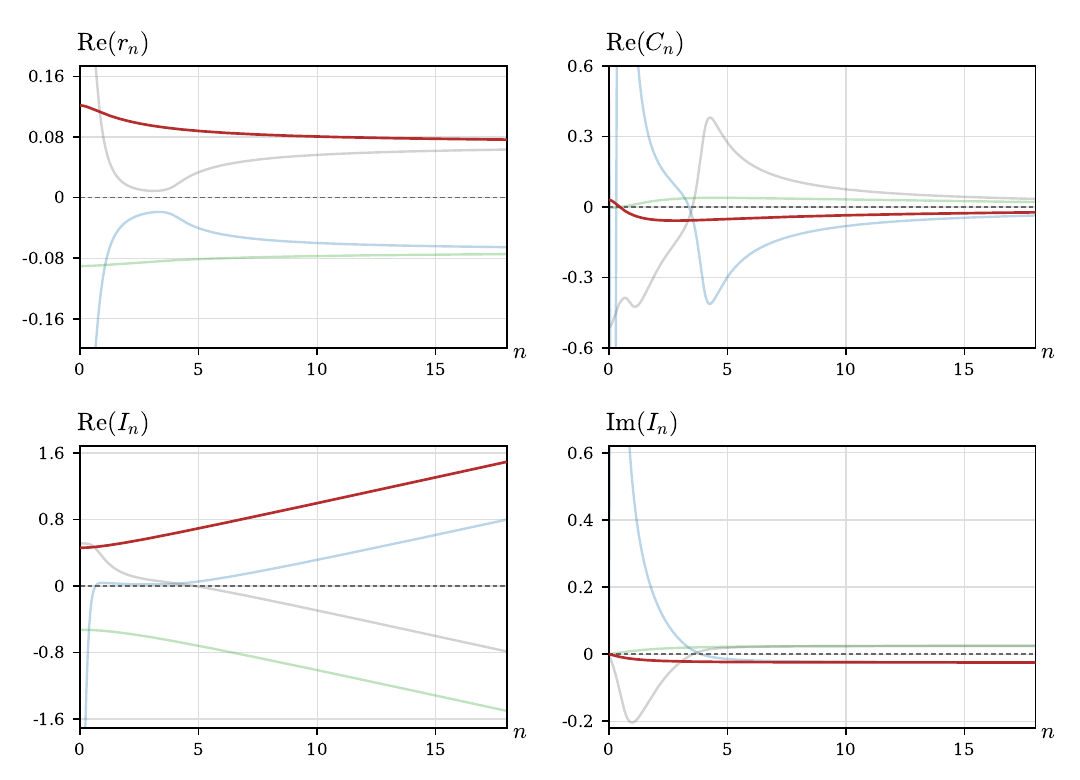}
    \vspace{-.7cm}
    \caption{The four branches of asymptotically flat Kerr-Newman saddles in the fixed angular velocity ensemble, with $\beta=5$, $Q=0.1$, and $\Omega_0=1$. For each integer shift $n$, the smoothness constraints \eqref{eq:app smoothness} determine complex saddle data $(r_n,a_n)$. In the upper row we plot the outer horizon $\mathrm{Re}(r_n)$ and the heat capacity $\mathrm{Re}(C_n)$, while in the lower row we plot the real and imaginary parts of the action $I_n$. The physical branch, with positive $r_0$, smaller action $I_0$, and positive heat capacity $C_0$ at $n=0$, corresponds to the small black hole and is shown in red. Its large $n$ behavior $\mathrm{Re}(I_n)\sim 2\pi Q^2 n$ shows convergence of the saddle sum along this branch. The large black hole, shown in grey, has larger $I_0$ and has $C_0<0$.}
    \label{fig:flatKN}
\end{figure}

\textbf{Kerr-Newman black hole ($Q\neq0$)}: We now study the case of a charged and rotating black hole. 
We solve for $a_n(\beta,\Omega_0,Q)$ and $r_n(\beta,\Omega_0,Q)$, for fixed $\beta$, $\Omega_0$, and $Q\neq0$ boundary conditions:
\begin{equation} \label{eq:app smoothness}
    \frac{a_n}{r_n^2+a_n^2} = \Omega_0 + \frac{4\pi\ii n}{\beta}\,,\quad \beta = \frac{4\pi r_n(r_n^2+a_n^2)}{r_n^2-a_n^2-Q^2} \,,\quad n\in\Z\,.
\end{equation}
In this case, there are four distinct families of solutions. A positive horizon coordinate $r_0(\beta,\Omega_0,Q)$ when $n=0$ is only possible for
\begin{equation}
    |\Omega_0 Q|<\frac{1}{2\sqrt2}\,,\quad \beta > \frac{6\sqrt3 \pi Q}{(1-8\Omega_0^2Q^2)\sqrt{1+\Omega_0^2Q^2}} \,,
\end{equation}
in which case two of the branches have positive $r_0(\beta,\Omega_0,Q)$. We will refer to the solution with smaller action $I_0$ and positive specific heat capacity $C_0=-\beta^2\p_\beta^2I_0$ as the physical branch, shown in red in Fig. \ref{fig:flatKN}. In asymptotically flat Kerr-Newman, this is the small black hole solution, while the grey curve in Fig. \ref{fig:flatKN} is the large black hole solution.

For large $n$, the four branches have solutions
\begin{equation}
\begin{split}
    &r_n = \sigma_r \frac{Q}{\sqrt2} + \frac1n \left(\sigma_a\frac{Q}{4\sqrt2} + \sigma_a\sigma_r\frac{\beta}{16\pi}\right) + \O\left(\frac{1}{n^2}\right) \\
    &a_n = \ii\sigma_a \frac{Q}{\sqrt2} + \frac\ii n \left(\sigma_r\frac{Q}{4\sqrt2} - \frac{\beta}{16\pi}\right) + \O\left(\frac{1}{n^2}\right)
\end{split}
\end{equation}
parametrized by $\sigma_a,\sigma_r\in\{\pm1\}$. The physical branch corresponds to $\sigma_a=\sigma_r=+1$. The large $n$ asymptotic behavior of the action for the four branches is given by
\begin{equation}
    I_{\rm KN}\left(\beta,\Omega_0+\frac{4\pi\ii}{\beta}n,Q\right) = \sigma_a\sigma_r2\pi Q^2 n + \sigma_r \frac12\beta\left(\sqrt2 Q - \sigma_a \ii Q^2\Omega_0\right) + \O\left(\frac1n\right)\,.
\end{equation}
Hence for asymptotically flat Kerr-Newman, the sum over saddles seems to converge for large $n$ for the family of solutions that corresponds to the physical branch.

\subsection{AdS$_4$ Kerr-Newman partition function}
We would now like to check the convergence of
\begin{equation}
    Z_{\rm KNAdS_4}(\beta,\Omega_0,Q) = \sum_{n\in\Z} e^{-I_{\rm KNAdS_4}\left(\beta,\Omega_0+\frac{4\pi\ii}{\beta}n,Q\right)} Z_{\rm KNAdS_4}^{(1)} \left(\beta,\Omega_0+\frac{4\pi\ii}{\beta}n,Q\right) \,,
\end{equation}
where the classical action of each saddle is given by Eq.~(33) of \cite{Caldarelli:1999xj},
\begin{equation}
    I_{\rm KNAdS_4}(\beta,\Omega_0,Q) = \frac{\beta}{4\Xi}\left[r_+\left(1-\frac{r_+^2+a^2}{\ell^2}\right) + \frac{a^2}{r_+} + q^2\left(\frac{1}{r_+} - \frac{2r_+}{r_+^2+a^2}\right)\right] \,,
\end{equation}
for $q=\Xi Q$ and $\Xi = (1-a^2/\ell^2)$.

\begin{figure}[h]
    \vspace{-.5cm}
    \includegraphics[width=\textwidth]{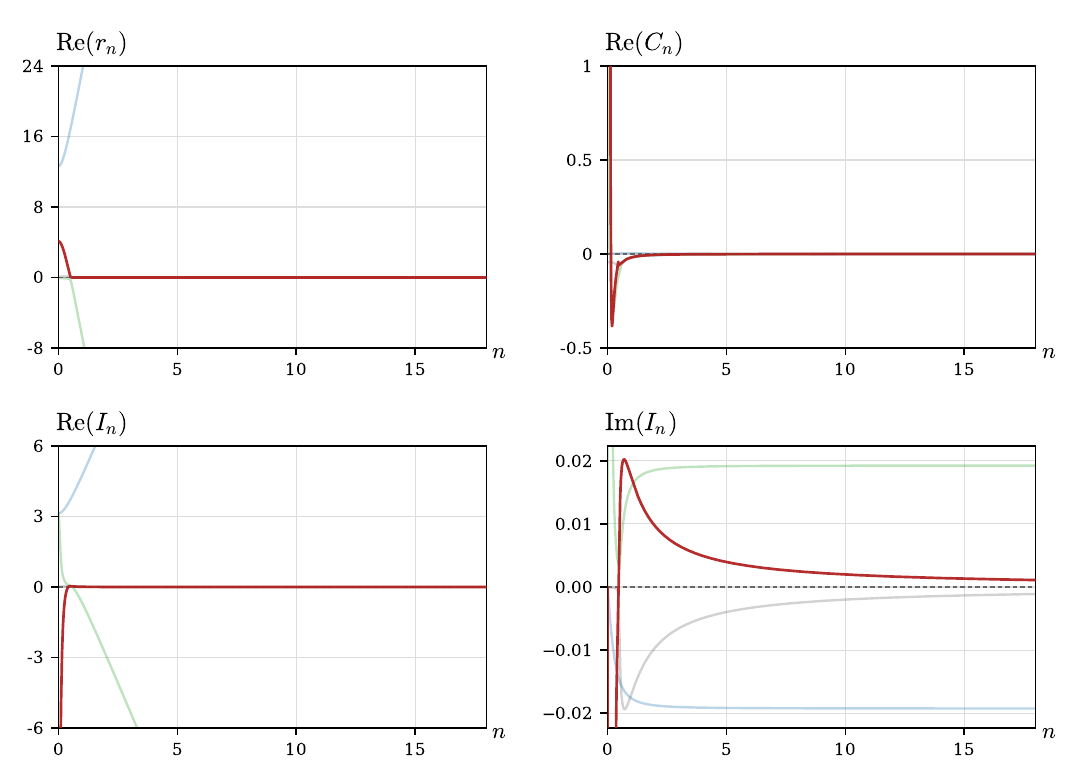}
    \vspace{-.7cm}
    \caption{The four branches of AdS Kerr saddles in the fixed angular velocity ensemble, with $\beta=1$, $Q=0$, $\Omega_0=0.1$, and $\ell=1$. For each integer shift $n$, the smoothness constraints \eqref{eq:app_ads4_kn} determine complex saddle data $(r_n,a_n)$. In the upper row we plot the outer horizon $\mathrm{Re}(r_n)$ and the heat capacity $\mathrm{Re}(C_n)$, while in the lower row we plot the real and imaginary parts of the action $I_n$. The physical branch, with positive $r_0$, $|a_0|<\ell$, smaller action $I_0$, and positive heat capacity $C_0$, corresponds to the large AdS black hole and is shown in red. The grey branch also has $r_0>0$ and $|a_0|<\ell$, but corresponds to the small AdS black hole and has $C_0<0$.}
    \label{fig:adsKerr}
\end{figure}

\textbf{Kerr AdS$_4$ black hole ($Q=0$)}: We study the case of a non-charged, rotating black hole in AdS$_4$, where the angular momentum parameter $a_n(\beta,\Omega_0)$ and outer horizon $r_+(\beta,\Omega_0)$ are solutions to the system of constraints given by \eqref{eq:ads4 omega periodicity}:
\begin{equation} \label{eq:app_ads4_kn}
    \frac{a_n(r_n^2 + \ell^2)}{\ell^2(r_n^2 + a_n^2)} = \Omega_0 + \frac{4\pi\ii n}{\beta} \,,\quad \beta = \frac{4\pi(r_n^2 + a_n^2)}{r_n\left(1 + \frac{1}{\ell^2} (a_n^2+3r_n^2) - \frac{a_n^2+Q^2(1-a_n^2/\ell^2)^2}{r_n^2}\right)} \,.
\end{equation}
for $Q=0$ and fixed $\beta$ and $\Omega_0$ boundary conditions. There are two distinct families of positive $r_0(\beta,\Omega_0)$ solutions, which exist when 
\begin{equation}
    \beta < \frac{8\pi\ell y^3}{3y^4+1}\,,\quad \Omega_0 = \frac{(1+y^2)^{3/2}\sqrt{3y^2-1}}{4\ell y^3} \,.
\end{equation}
These correspond to the usual small and large black holes in AdS, each of which has two branches of $a_0(\beta,\Omega_0)$. We will see that the thermodynamically stable black holes correspond to the large black hole with $|a_0|<\ell$. 

There are two branches with large $n$ asymptotic solutions
\begin{equation}
\begin{split}
    r_n &= \sigma_1\frac{8\pi\ell^2}{\sqrt3\beta}n + \left(\frac{8\pi\ell^2}{3\beta} - \sigma_1\frac{2\ii\ell^2\Omega_0}{\sqrt3}\right) + \frac{\sigma_1}{n} \frac{3\beta^2+8\pi^2\ell^2}{24\sqrt3\pi\beta} + \O\left(\frac{1}{n^2}\right) \\
    a_n &= -\frac{8\pi\ii\ell^2}{\beta}n - \left(2\ell^2\Omega_0 + \sigma_1\frac{4\sqrt3\ii\pi\ell^2}{\beta}\right) - \frac{\ii}{n} \frac{3\beta^2+16\pi^2\ell^2}{8\pi\beta} + \O\left(\frac{1}{n^2}\right)
\end{split}
\end{equation}
and action
\begin{equation}
    I_{\rm KAdS_4}\left(\beta,\Omega_0+\frac{4\pi\ii}{\beta}n\right) = \sigma_1\frac{4\ell^2\pi}{3\sqrt3}n + \frac{\ell^2(6\pi-\ii\sigma_1\sqrt3\beta\Omega_0)}{9} + \O\left(\frac1n\right)\,,
\end{equation}
where $\sigma_1=\pm1$. The other two branches have large $n$ behavior
\begin{equation}
\begin{split}
    r_n &= -\ii\frac{\sigma_2}{n}\frac{\beta}{8\pi} + \frac{\sigma_2}{n^2}\frac{\beta^2\Omega_0}{32\pi^2} + \O\left(\frac{1}{n^3}\right) \\
    a_n &= - \ii\frac{1}{n}\frac{\beta}{8\pi} + \frac{1}{n^2}\frac{2\pi\beta\sigma_2+\beta^2\Omega_0}{32\pi^2} + \O\left(\frac{1}{n^3}\right)
\end{split}
\end{equation}
and action
\begin{equation}
    I_{\rm KAdS_4}\left(\beta,\Omega_0+\frac{4\pi\ii}{\beta}n\right) = -\ii\frac{\sigma_2}{n} \frac{\beta^2}{16\pi} + \frac{1}{n^2} \frac{\beta^2(2\pi+\sigma_2\beta\Omega_0)}{64\pi^2} + \O\left(\frac{1}{n^3}\right)\,,
\end{equation}
where $\sigma_2=\pm1$.

The physical branch, shown in red in Fig. \ref{fig:adsKerr}, is of the second type, with $\sigma_2=-1$. Since the action vanishes for large $n$, we cannot conclude whether the sum converges or diverges without computing the one-loop determinant.

\begin{figure}[h]
    \vspace{-.5cm}
    \includegraphics[width=\textwidth]{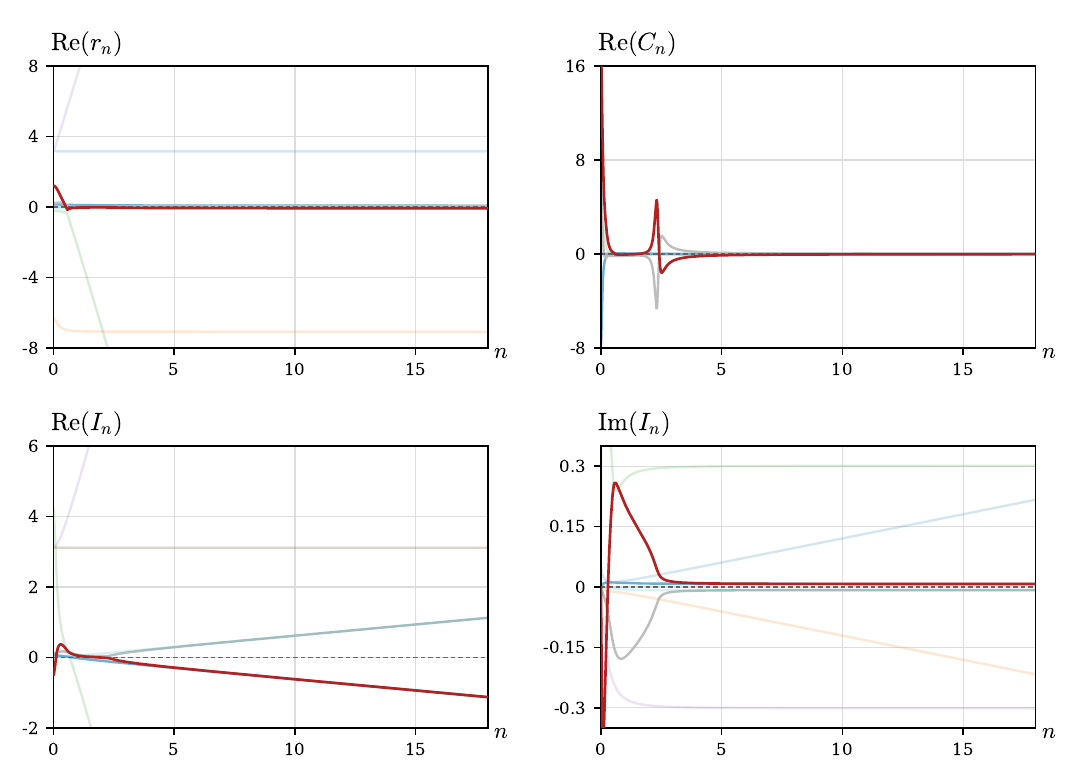}
    \vspace{-.7cm}
    \caption{The eight branches of AdS Kerr-Newman saddles in the fixed angular velocity ensemble, with $(r_n,a_n)$ determined from the smoothness constraints \eqref{eq:app_ads4_kn} using $\beta=3$, $Q=0.1$, $\Omega_0=0.5$, and $\ell=1$. In the upper row we plot the outer horizon $\mathrm{Re}(r_n)$ and the heat capacity $\mathrm{Re}(C_n)$, while in the lower row we plot the real and imaginary parts of the action $I_n$. The physical branch, with positive $r_0$, $|a_0|<\ell$, smaller action $I_0$, and positive heat capacity $C_0$, corresponds to the large AdS black hole and is shown in red. The grey and darker blue branches also have $r_0>0$ and $|a_0|<\ell$, but correspond to small AdS black holes. The physical branch has action growing as $-n$, and all eight branches have either real or imaginary part of the action growing with $\pm n$.}
    \label{fig:adskn}
\end{figure}

\textbf{Kerr-Newman AdS$_4$ black hole ($Q\neq0$)}: There are now eight solutions to the system of constraints \eqref{eq:ads4 omega periodicity}, shown in Fig. \ref{fig:adskn}. Four of the branches have large $n$ asymptotics
\begin{equation}
\begin{split}
    r_n &= \ii\sigma_1\frac{4 k\ell^2\pi}{\sqrt{k-1}\beta}n + \left(\frac{2(2-k)\ell^2\pi}{(5+k)\beta}+\sigma_1\frac{k\ell^2\Omega_0}{\sqrt{k-1}}\right) + \O\left(\frac{1}{n}\right) \\
    a_n &= \ii\frac{4k\ell^2\pi}{\beta}n + \left(k\ell^2\Omega_0 - \sigma_1\frac{4(k-1)^{3/2}\ell^2\pi}{(k+5)\beta}\right) + \O\left(\frac{1}{n}\right)
\end{split}
\end{equation}
and action
\begin{equation}
    I_{\rm KNAdS_4}\left(\beta,\Omega_0+\frac{4\pi\ii}{\beta}n\right) = \ii\sigma_1\frac{4\pi\ell^2}{(k-1)^{3/2}}n + \O\left(\frac{1}{n^0}\right)\,,
\end{equation}
where $k=\frac{1+2Q^2/\ell^2+\sigma_2\sqrt{1+12Q^2/\ell^2}}{2Q^2/\ell^2}$ and $(\sigma_1,\sigma_2)=(\pm1,\pm1)$. The other four black hole branches have large $n$ behavior
\begin{equation}
\begin{split}
    r_n &= \frac{Q\sigma_3}{\sqrt{2-Q^2/\ell^2}} + \O\left(\frac{1}{n}\right) \\
    a_n &= \ii\frac{Q\sigma_3\sigma_4}{\sqrt{2-Q^2/\ell^2}} + \O\left(\frac{1}{n}\right)
\end{split}
\end{equation}
and action
\begin{equation}
    I_{\rm KNAdS_4}\left(\beta,\Omega_0+\frac{4\pi\ii}{\beta}n\right) = -2\pi\sigma_4 Q^2 n + \left(\ii\sigma_4 \frac{\beta Q^2\Omega_0}{2} + \sigma_3 \frac{\beta Q^3}{2\ell^2\sqrt{2-Q^2/\ell^2}}\right) + \O\left(\frac{1}{n}\right)\,,
\end{equation}
where $(\sigma_3,\sigma_4)=(\pm1,\pm1)$. The large $n$ asymptotics of the real and imaginary parts of the action are shown in the bottom row of Fig. \ref{fig:adskn}, from which we can see that in this case, the physical branch corresponds to $(\sigma_3,\sigma_4)=(-1,+1)$. The action of the physical branch for the choice of parameters in Fig. \ref{fig:adskn} grows as $I\sim -2\pi Q^2n$, such that the partition function diverges for large $n$. However, for different choices of $(\beta,Q,\Omega_0)$, the action can also grow with positive $n$, so there is no simple answer to the convergence of the partition function for AdS Kerr-Newman.

\section{The sum over saddles in the near-extremal limit} \label{app:variance}

In this appendix, we study the sum over fixed $(\beta,\Omega_0,Q)$ saddles in the near-extremal limit $\beta\gg Q$ of asymptotically flat Kerr-Newman black holes. Recall that $\Omega_0$ is an angular velocity obtained from a Wick rotation of a Lorentzian rotating solution. We will separately consider the cases of $\Omega_0\in\R$, corresponding to a complex Euclidean metric coming from a real Lorentzian angular velocity, and $\Omega_0\in\ii\R$, corresponding to a real Euclidean metric with a real Euclidean angular velocity.

\subsection{Real Lorentzian angular velocity}
In the near-extremal limit $\beta\gg Q$, the on-shell action for the family of classical Kerr-Newman saddles given by \eqref{eq:4dkerrnewman} has the form
\begin{equation} \label{eq:appKNeps}
    -I_{\rm KN}\left(\beta,\Omega_n,Q\right) = \frac{\beta Q}{G_N}f(\epsilon_n) + \frac{Q^2}{G_N}g(\epsilon_n) + \O\qty(\frac{Q^3}{\beta G_N})
\end{equation}
where $\epsilon\equiv Q\Omega_0$ and $\epsilon_n\equiv Q\Omega_n=\epsilon+\frac{4\pi\ii Qn}{\beta}$ for dimensionless $\epsilon$ and $\epsilon_n$, and we have defined
\begin{equation}
    f(\epsilon_n) = -\frac{\qty(3+\sqrt{1-8\epsilon_n^2})^{3/2}}{4\sqrt{2\qty(1+\sqrt{1-8\epsilon_n^2})}}\,, \quad g(\epsilon_n) = \frac{2\pi}{1+\sqrt{1-8\epsilon_n^2}}\,.
\end{equation}
For a typical Kerr-Newman black hole, note that $\epsilon = \O(1)$ and $f(\epsilon) = \O(1)$. To understand the $n$-dependence of the action, it is useful to expand \eqref{eq:appKNeps} in a power series in $\frac{nQ}{\beta}$. To do this, let
\begin{equation} \label{eq:app_Jcl}
    J_{z,\text{cl}} \equiv -\frac{1}{\beta}\p_{\Omega_0}I_{\rm KN}(\beta,\Omega_0,Q) = \frac{Q^2}{G_N}\qty[f'(\epsilon) + \frac{Q}{\beta}g'(\epsilon)+\ldots]
\end{equation}
and\footnote{This is the standard definition of the variance if the Lorentzian angular velocity is real. If the angular velocity is complex, then the definition Eq. \eqref{eq:app_var} still holds, but $\mathrm{var}(J_{z,\text{cl}})$ can in principle be complex.}
\begin{equation} \label{eq:app_var}
    \mathrm{var}(J_{z,\text{cl}}) \equiv -\frac{1}{\beta^2}\p_{\Omega_0}^2I_{\rm KN}(\beta,\Omega_0,Q) = \frac{Q^3}{\beta G_N}\qty[f''(\epsilon) + \frac{Q}{\beta}g''(\epsilon)+\ldots]
\end{equation}
such that the action \eqref{eq:appKNeps} takes the form
\begin{equation} \label{eq:app_ndep}
    -I_{\rm KN}\qty(\beta,\Omega_0+\frac{4\pi\ii n}{\beta},Q) = -I_{\rm KN}(\beta,\Omega_0,Q) + 4\pi\ii J_{z,\text{cl}}n - 8\pi^2\mathrm{var}(J_{z,\text{cl}})n^2 + \O(n^3)\,.
\end{equation}
The order $n^3$ term is suppressed relative to the quadratic term by $\frac{Q}{\beta}$, so the Gaussian approximation is controlled in the near-extremal limit $\beta\gg Q$. Higher order terms are further suppressed in powers of $\frac{Q}{\beta}$. Thus, to a good approximation the action \eqref{eq:app_ndep} is quadratic in $n$. At this point, we will further separate two cases: a semiclassical regime $16\pi^2\mathrm{var}(J_{z,\text{cl}})\gg1$, in which the $n$-dependence of \eqref{eq:app_ndep} causes one image to dominate, and a quantum regime $16\pi^2\mathrm{var}(J_{z,\text{cl}})\ll1$, in which the $n$-dependence of \eqref{eq:app_ndep} is suppressed at low temperatures so the images all contribute at the same order.\footnote{This should be contrasted with the 3D BTZ case, in which, even in the near-extremal limit, the $n$ dependence is not suppressed in $\beta$, so one saddle still dominates the partition function.}

\subsubsection*{Semiclassical regime near extremality: $\beta\gg Q$ and  $\beta G_N\ll 16\pi^2Q^3$}
When the image label $n$ has a narrow distribution ($\frac{1}{16\pi^2\mathrm{var}(J_{z,\text{cl}})}\ll1$), one integer image dominates:
\begin{equation}
    n_* = 0 \,.
\end{equation}
The magnitude of the $n$th image relative to the $n=0$ image is exponentially suppressed,
\begin{equation}
    \left|\frac{e^{-I_{\rm KN}\left(\beta,\Omega_0+\frac{4\pi\ii n}{\beta},Q\right)}}{e^{-I_{\rm KN}\left(\beta,\Omega_0,Q\right)}}\right| \approx e^{-8\pi^2\mathrm{var}(J_{z,\text{cl}})n^2} \,,
\end{equation}
such that the partition function is well-approximated by
\begin{equation}
    Z = \sum_{n\in\Z} e^{-I_{\rm KN}\left(\beta,\Omega_0+\frac{4\pi\ii n}{\beta},Q\right)} \approx e^{-I_{\rm KN}(\beta,\Omega_0,Q)}\,.
\end{equation}
The thermodynamic angular momentum
\begin{equation} \label{eq:appJ_th}
    \braket{J_z} \equiv \frac{1}{\beta}\p_{\Omega_0}\log Z \approx \frac{Q^2}{G_N}f'(\epsilon)
\end{equation}
matches the classical angular momentum \eqref{eq:app_Jcl} associated to the single dominant saddle, and the variance
\begin{equation} \label{app:varJ_th}
    \mathrm{var}(J_z) \equiv \frac{1}{\beta^2}\p_{\Omega_0}^2\log Z \approx \frac{Q^3}{\beta G_N} f''(\epsilon)
\end{equation}
also matches the variance of the classical angular momentum \eqref{eq:app_var}.\footnote{One can verify that the average angular momentum is positive (negative) for $\epsilon>0$ ($\epsilon<0$), while the variance is always positive.} In the limit $\beta G_N\ll Q^3$, the thermodynamic variance is large, so many discrete spin values contribute inside the dominant $n_*=0$ image saddle. In this sense, $J_z$ may be treated as a continuous variable.

\subsubsection*{The quantum regime near extremality: $\beta G_N\gg 16\pi^2Q^3$}
When the image label $n$ has a broad distribution ($\frac{1}{16\pi^2\mathrm{var}(J_{z,\text{cl}})}\gg1$), many $n$-images contribute comparably to the partition function. In such a case the $n$-dependence of the one-loop determinant is in principle important -- however, since the $n$-dependence depends on the exact details about the theory and the black hole (e.g., whether the black hole is supersymmetric or not and whether we are working in a theory of supergravity or not) we will neglect the one-loop determinant in the calculation below. Consider the Poisson resummation of the near-extremal Kerr-Newman partition function
\begin{equation}
    Z \approx \sum_{n\in\Z} e^{-I_{\rm KN}(\beta,\Omega_0,Q) + 4\pi\ii nJ_{z,\text{cl}} - 8\pi^2\mathrm{var}(J_{z,\text{cl}})n^2} = \frac{e^{-I_{\rm KN}(\beta,\Omega_0,Q)}}{\sqrt{8\pi\,\mathrm{var}(J_{z,\text{cl}})}} \sum_{J_z\in\frac12\Z} e^{-\frac{1}{2\,\mathrm{var}(J_{z,\text{cl}})}[J_z-J_{z,\text{cl}}]^2} \,.
\end{equation}
The thermodynamic angular momentum and variance are given by
\begin{equation*}
\begin{split}
    \braket{J_z} &\equiv \frac{1}{\beta}\p_{\Omega_0}\log Z \approx J_{z,\text{cl}} + \frac{Q}{\beta} \p_{\epsilon}\log\left(\qty[\mathrm{var}(J_{z,\text{cl}})]^{-1/2}\sum_{J_z\in\frac12\Z} e^{-\frac{1}{2\,\mathrm{var}(J_{z,\text{cl}})}(J_z-J_{z,\text{cl}})^2}\right) \,,\\
    \mathrm{var}(J_z) &\equiv \frac{1}{\beta^2}\p_{\Omega_0}^2\log Z \approx \mathrm{var}(J_{z,\text{cl}}) + \frac{Q^2}{\beta^2} \p_{\epsilon}^2\log\left(\qty[\mathrm{var}(J_{z,\text{cl}})]^{-1/2}\sum_{J_z\in\frac12\Z} e^{-\frac{1}{2\,\mathrm{var}(J_{z,\text{cl}})}(J_z-J_{z,\text{cl}})^2}\right) \,,
\end{split}
\end{equation*}
which we compute using the Poisson resummed partition function.\footnote{Note that before Poisson resummation, the variance of $J_z$ is computed by
\begin{equation}
    \mathrm{var}(J_z) \equiv \frac{1}{\beta^2}\p_{\Omega_0}^2\log Z = \frac{1}{Z}\sum_n (j_n^2 + \chi_n)e^{-I_n} - \left(\frac{1}{Z}\sum_n j_n e^{-I_n}\right)^2
\end{equation}
where $I_n\equiv I_{\rm KN}\qty(\beta,\Omega_0+\frac{4\pi\ii n}{\beta},Q)$, $j_n\equiv-\frac{1}{\beta}\p_{\Omega_0}I_n$, and $\chi_n\equiv-\frac{1}{\beta^2}\p_{\Omega_0}^2I_n$, while after Poisson resummation it is computed by
\begin{equation}
    \mathrm{var}(J_z) = \frac{\sum_{J_z\in\frac12\Z} J_z^2 e^{-\frac{1}{2\,\mathrm{var}(J_{z,\text{cl}})}[J_z-J_{z,\text{cl}}]^2}}{\sum_{J_z\in\frac12\Z} e^{-\frac{1}{2\,\mathrm{var}(J_{z,\text{cl}})}[J_z-J_{z,\text{cl}}]^2}} - \left(\frac{\sum_{J_z\in\frac12\Z} J_z e^{-\frac{1}{2\,\mathrm{var}(J_{z,\text{cl}})}[J_z-J_{z,\text{cl}}]^2}}{\sum_{J_z\in\frac12\Z} e^{-\frac{1}{2\,\mathrm{var}(J_{z,\text{cl}})}[J_z-J_{z,\text{cl}}]^2}}\right)^2 \,.
\end{equation}}
In the semiclassical regime described above, many $J_z$ lie under the Gaussian, such that the sum is approximately independent of $J_{z,\text{cl}}$ (and hence of $\Omega_0$) and we recover \eqref{eq:appJ_th}. On the other hand, in the quantum regime we find, to leading order,
\begin{equation}
    \braket{J_z} = \frac{\sum_{J_z\in\frac12\Z} J_z e^{-\frac{1}{2\,\mathrm{var}(J_{z,\text{cl}})}[J_z-J_{z,\text{cl}}]^2}}{\sum_{J_z\in\frac12\Z} e^{-\frac{1}{2\,\mathrm{var}(J_{z,\text{cl}})}[J_z-J_{z,\text{cl}}]^2}} \approx J_*\in\frac12\Z \,,
\end{equation}
where the sum is dominated by the half-integer $J_*$ closest to $J_{z,\text{cl}}$, for $\delta\equiv |J_{z,\text{cl}}-J_*|\in\left(0,\frac14\right)$. The variance is given by
\begin{equation} \label{eq:app_variance}
    \mathrm{var}(J_z) \approx \frac14 e^{-\frac{\beta G_N}{2Q^3f''(\epsilon)}\qty(\frac14-\delta)} \,,
\end{equation}
which is exponentially suppressed in the quantum regime $\beta G_N\gg 16\pi^2Q^3$.\footnote{For the corner cases of $\delta=0$ and $\delta=\frac14$, the variance of $J_{z,\text{cl}}$ has to be studied separately. Then the classical angular momentum is precisely in between two half-integers: $J_{z,1}+\frac14=J_{z,\text{cl}}=J_{z,2}-\frac14$, and one can show that
\begin{equation}
    \braket{J_z} \approx J_{z,\text{cl}}\,,\qquad \mathrm{var}(J_z)\big|_{\delta=\frac14} \approx \frac{1}{16}\,,\qquad \mathrm{var}(J_z)\big|_{\delta=0} \approx \frac12 e^{-\frac{\beta G_N}{8Q^3f''(\epsilon)}}\,.
\end{equation}
In the case $\delta=0$, $J_*=J_{z,\text{cl}}$ dominates the path integral, while in the case $\delta=\frac14$ both $J_{z,1}$ and $J_{z,2}$ contribute equally to the path integral.}

Note that in this regime, the semiclassical partition function, which includes a single $n$ saddle, is not a good approximation: many $n$-images contribute to the partition function, while the angular momentum $J_z$ is sharply peaked. Whereas approximating the partition function by a single saddle would have resulted in a polynomially suppressed variance $\mathrm{var}(J_z) \sim \frac{Q^3}{\beta G_N}$, the variance computed using the full partition function is generically exponentially suppressed in $\frac{\beta G_N}{Q^3}$.

\subsection{Real Euclidean angular velocity}
Consider now the analytic continuation $\Omega_{n,e} = \ii\Omega_n = \Omega_e-\frac{4\pi n}{\beta}$ where $\Omega_e=\ii\Omega_0$, such that the saddles are real in Euclidean signature. Let $\theta=\beta\Omega_e$ and $\theta_n = \beta\Omega_{n,e} = \theta-4\pi n$. Note that due to the periodicity we can define $\theta \in [0, 4\pi)$. In the near-extremal limit $\beta\gg Q$, the on-shell action for the family of classical Kerr-Newman saddles given by \eqref{eq:4dkerrnewman} now takes the form
\begin{equation} \label{eq:appKNeps_I}
    -I_{\rm KN}(\beta,\theta-4\pi n,Q) \approx -I_{\rm RN} - 8\pi^2\mathrm{var}_e(J_{z,\text{cl}})\qty(\frac{\theta}{4\pi}-n)^2 \,,
\end{equation}
where $I_{\rm RN}$ is the action for the Reissner-Nordstr\"om black hole, given by evaluating $I_{\rm KN}$ at $\Omega_e=0$, and we have defined
\begin{equation}
    \mathrm{var}_e(J_{z,\text{cl}}) \equiv \p_{\theta}^2I_{\rm KN}(\beta,\theta,Q) = \frac{Q^3}{\beta G_N}\left(1 + \frac{4\pi Q}{\beta} + \ldots\right) \,.
\end{equation}

\subsubsection*{Semiclassical regime near extremality: $\beta\gg Q$ and $\beta G_N\ll 16\pi^2Q^3$}

In this case, $\frac{1}{16\pi^2\mathrm{var}_e(J_{z,\text{cl}})}\ll1$ and the image that dominates the partition function corresponds to the integer $n$ that minimizes \eqref{eq:appKNeps_I}, given by
\begin{equation}
    n_* = \text{nearest integer to } \qty(\frac{\theta}{4\pi}) \,.
\end{equation}
The partition function is approximated by
\begin{equation}
    Z = \sum_ne^{-I_{\rm KN}(\beta,\theta-4\pi n,Q)} \approx e^{-I_{\rm RN}} e^{-8\pi^2\mathrm{var}_e(J_{z,\text{cl}})\qty(\frac{\theta}{4\pi}-n_*)^2}\,,
\end{equation}
and the thermodynamic angular momentum by
\begin{equation}
    \braket{J_z} \equiv \ii\p_{\theta}\log Z \approx \ii\frac{4\pi Q^3}{\beta G_N} \qty(n_* - \frac{\theta}{4\pi}) \,,
\end{equation}
where we note that $|n_* - \frac{\theta}{4\pi}| \in \left[0,\frac12\right)$.\footnote{Here the corner case is $\frac{\theta}{4\pi}\in\Z+\frac12$. For such values of $\theta$, bosonic and fermionic black hole states are weighed by a relative minus sign. Then the partition function is approximated by contributions from both $n_1=\frac{\theta}{4\pi}+\frac12$ and $n_2=\frac{\theta}{4\pi}-\frac12$, and the average angular momentum is given by $\braket{J_z}\approx 0$. The variance is still large in this case.} Similarly to \eqref{app:varJ_th}, the thermodynamic variance is large so $J_z$ is effectively continuous.

\subsubsection*{Quantum regime near extremality: $\beta G_N\gg 16\pi^2Q^3$}
For $\frac{1}{16\pi^2\mathrm{var}_e(J_{z,\text{cl}})}\gg1$, consider the Poisson resummation of the near-extremal Kerr-Newman partition function, given by
\begin{equation}
    Z = \frac{1}{\sqrt{8\pi\,\mathrm{var}_e(J_{z,\text{cl}})}} e^{-I_{\rm RN}} \sum_{J_z\in\frac12\Z} e^{-\frac{1}{2\,\mathrm{var}_e(J_{z,\text{cl}})}J_z^2} e^{-\ii\theta J_z}
\end{equation}
The sum over $J_z$ is dominated by $J_z=0$,\footnote{Alternatively, one can approximate the sum over $n$ by a Gaussian integral
\begin{equation}
    Z = \sum_ne^{-I_{\rm KN}\left(\beta,\Omega_e-\frac{4\pi n}{\beta},Q\right)} \approx e^{-I_{\rm RN}} \int_{-\infty}^\infty dn e^{-8\pi^2\mathrm{var}_e(J_{z,\text{cl}})\qty(\frac{\theta}{4\pi}-n)^2} = e^{-I_{\rm RN}} \frac{1}{\sqrt{8\pi\,\mathrm{var}_e(J_{z,\text{cl}})}}
\end{equation}
which is a reasonable approximation for $\beta G_N\gg 16\pi^2Q^3$, i.e. for the regime in which many images lie under the Gaussian. A shift by $\theta$ only shifts the center of the integral, so $Z$ is approximately independent of $\theta$ in this regime, and hence $\braket{J_z}\approx0$.}
\begin{equation}
    \braket{J_z} \equiv \ii\p_\theta\log Z = \frac{\sum_{J_z\in\frac12\Z} J_ze^{-\frac{1}{2\,\mathrm{var}_e(J_{z,\text{cl}})}J_z^2} e^{-\ii\theta J_z}}{\sum_{J_z\in\frac12\Z} e^{-\frac{1}{2\,\mathrm{var}_e(J_{z,\text{cl}})}J_z^2} e^{-\ii\theta J_z}} \approx 0\,,
\end{equation}
and just as in \eqref{eq:app_variance}, the variance is generically exponentially suppressed in $\frac{\beta G_N}{Q^3}$.

There is a simple way to understand why the spin distribution is centered at $J_z=0$ near extremality. At fixed charge $Q$, the Kerr-Newman extremality bound is 
\begin{equation}
    M\geq M_{\text{ext}} = \frac{1}{G_N}\sqrt{\frac{Q^2+\sqrt{Q^4+4 G_N^2J_z^2}}{2}} \approx \frac{Q}{G_N} + \frac{G_NJ_z^2}{2Q^3} + \ldots \,.
\end{equation}
Thus $M_{\text{ext}}$ is an even function of $J_z$ with its minimum at $J_z=0$. Near extremality, nonzero spin raises the minimum allowed mass, so at fixed $Q$ the extremal bound is easiest to satisfy at $J_z=0$.

\newpage
\bibliographystyle{JHEP}
\bibliography{references}

@article{Kolanowski:2026gii,
    author = "Kolanowski, Maciej and Marolf, Donald",
    title = "{How to tame your (black hole) saddles: Lessons from the Lorentzian Gravitational Path Integral}",
    eprint = "2603.24681",
    archivePrefix = "arXiv",
    primaryClass = "hep-th",
    month = "3",
    year = "2026"
}

@article{Boruch:2022tno,
    author = "Boruch, Jan and Heydeman, Matthew T. and Iliesiu, Luca V. and Turiaci, Gustavo J.",
    title = "{BPS and near-BPS black holes in AdS$_{5}$ and their spectrum in $ \mathcal{N} $ = 4 SYM}",
    eprint = "2203.01331",
    archivePrefix = "arXiv",
    primaryClass = "hep-th",
    doi = "10.1007/JHEP07(2025)220",
    journal = "JHEP",
    volume = "07",
    pages = "220",
    year = "2025"
}

@article{Cabo-Bizet:2025kjc,
    author = "Cabo-Bizet, Alejandro",
    title = "{Why Indices Count the Total Number of Black Hole Microstates (at large N)}",
    eprint = "2512.19946",
    archivePrefix = "arXiv",
    primaryClass = "hep-th",
    month = "12",
    year = "2025"
}

@article{Migdal:1975zg,
    author = "Migdal, Alexander A.",
    editor = "Khalatnikov, I. M. and Mineev, V. P.",
    title = "{Recursion equations in gauge field theories}",
    reportNumber = "PRINT-75-1043 (LANDAU-INST)",
    journal = "Sov. Phys. JETP",
    volume = "42",
    pages = "413--418",
    year = "1975"
}

@article{Coleman:1991ku,
    author = "Coleman, Sidney R. and Preskill, John and Wilczek, Frank",
    title = "{Quantum hair on black holes}",
    eprint = "hep-th/9201059",
    archivePrefix = "arXiv",
    reportNumber = "IASSNS-HEP-91-64, CALT-68-1764, HUTP-92-A003",
    doi = "10.1016/0550-3213(92)90008-Y",
    journal = "Nucl. Phys. B",
    volume = "378",
    pages = "175--246",
    year = "1992"
}

@article{Singhi:2025rfy,
    author = "Singhi, Kaustubh",
    title = "{Complex Kerr-AdS black holes}",
    eprint = "2510.01313",
    archivePrefix = "arXiv",
    primaryClass = "hep-th",
    doi = "10.1007/JHEP02(2026)138",
    journal = "JHEP",
    volume = "02",
    pages = "138",
    year = "2026"
}

@article{Mahajan:2025bzo,
    author = "Mahajan, Raghu and Singhi, Kaustubh",
    title = "{A brief note on complex AdS-Schwarzschild black holes}",
    eprint = "2509.08883",
    archivePrefix = "arXiv",
    primaryClass = "hep-th",
    doi = "10.1007/JHEP11(2025)164",
    journal = "JHEP",
    volume = "11",
    pages = "164",
    year = "2025"
}

@article{Boruch:2026hbr,
    author = "Boruch, Jan and Tabor, Elisa and Turiaci, Gustavo J.",
    title = "{3D Gravity and Chaos in CFTs with Fermions}",
    eprint = "2602.17618",
    archivePrefix = "arXiv",
    primaryClass = "hep-th",
    month = "2",
    year = "2026"
}

@article{Gibbons:2004ai,
    author = "Gibbons, G. W. and Perry, M. J. and Pope, C. N.",
    title = "{The First law of thermodynamics for Kerr-anti-de Sitter black holes}",
    eprint = "hep-th/0408217",
    archivePrefix = "arXiv",
    reportNumber = "DAMTP-2004-87, MIFP-04-17",
    doi = "10.1088/0264-9381/22/9/002",
    journal = "Class. Quant. Grav.",
    volume = "22",
    pages = "1503--1526",
    year = "2005"
}

@article{Giombi:2008vd,
    author = "Giombi, Simone and Maloney, Alexander and Yin, Xi",
    title = "{One-loop Partition Functions of 3D Gravity}",
    eprint = "0804.1773",
    archivePrefix = "arXiv",
    primaryClass = "hep-th",
    doi = "10.1088/1126-6708/2008/08/007",
    journal = "JHEP",
    volume = "08",
    pages = "007",
    year = "2008"
}

@article{Banados:1992wn,
    author = "Banados, Maximo and Teitelboim, Claudio and Zanelli, Jorge",
    title = "{The Black hole in three-dimensional space-time}",
    eprint = "hep-th/9204099",
    archivePrefix = "arXiv",
    reportNumber = "PRINT-92-0151 (CHILE), IASSNS-HEP-92-29",
    doi = "10.1103/PhysRevLett.69.1849",
    journal = "Phys. Rev. Lett.",
    volume = "69",
    pages = "1849--1851",
    year = "1992"
}

@article{Witten:1991we,
    author = "Witten, Edward",
    title = "{On quantum gauge theories in two-dimensions}",
    doi = "10.1007/BF02100009",
    journal = "Commun. Math. Phys.",
    volume = "141",
    pages = "153--209",
    year = "1991"
}

@article{Witten:1992xu,
    author = "Witten, Edward",
    title = "{Two-dimensional gauge theories revisited}",
    eprint = "hep-th/9204083",
    archivePrefix = "arXiv",
    doi = "10.1016/0393-0440(92)90034-X",
    journal = "J. Geom. Phys.",
    volume = "9",
    pages = "303--368",
    year = "1992"
}

@article{Stanford:2017thb,
    author = "Stanford, Douglas and Witten, Edward",
    title = "{Fermionic Localization of the Schwarzian Theory}",
    eprint = "1703.04612",
    archivePrefix = "arXiv",
    primaryClass = "hep-th",
    doi = "10.1007/JHEP10(2017)008",
    journal = "JHEP",
    volume = "10",
    pages = "008",
    year = "2017"
}

@article{Iliesiu:2022kny,
    author = "Iliesiu, Luca V. and Murthy, Sameer and Turiaci, Gustavo J.",
    title = "{Black hole microstate counting from the gravitational path integral}",
    eprint = "2209.13602",
    archivePrefix = "arXiv",
    primaryClass = "hep-th",
    doi = "10.1007/JHEP08(2025)152",
    journal = "JHEP",
    volume = "08",
    pages = "152",
    year = "2025"
}

@article{Smarr:1972kt,
    author = "Smarr, Larry",
    title = "{Mass formula for Kerr black holes}",
    doi = "10.1103/PhysRevLett.30.71",
    journal = "Phys. Rev. Lett.",
    volume = "30",
    pages = "71--73",
    year = "1973",
    note = "[Erratum: Phys.Rev.Lett. 30, 521--521 (1973)]"
}

@article{Hawking:1995fd,
    author = "Hawking, S. W. and Horowitz, Gary T.",
    title = "{The Gravitational Hamiltonian, action, entropy and surface terms}",
    eprint = "gr-qc/9501014",
    archivePrefix = "arXiv",
    reportNumber = "DAMTP-R-94-52, UCSBTH-94-37",
    doi = "10.1088/0264-9381/13/6/017",
    journal = "Class. Quant. Grav.",
    volume = "13",
    pages = "1487--1498",
    year = "1996"
}

@article{Maldacena:1998bw,
    author = "Maldacena, Juan Martin and Strominger, Andrew",
    title = "{AdS(3) black holes and a stringy exclusion principle}",
    eprint = "hep-th/9804085",
    archivePrefix = "arXiv",
    reportNumber = "HUTP-98-A016",
    doi = "10.1088/1126-6708/1998/12/005",
    journal = "JHEP",
    volume = "12",
    pages = "005",
    year = "1998"
}

@article{Chen:2023mbc,
    author = "Chen, Yiming and Turiaci, Gustavo J.",
    title = "{Spin-statistics for black hole microstates}",
    eprint = "2309.03478",
    archivePrefix = "arXiv",
    primaryClass = "hep-th",
    doi = "10.1007/JHEP04(2024)135",
    journal = "JHEP",
    volume = "04",
    pages = "135",
    year = "2024"
}

@article{Maloney:2007ud,
    author = "Maloney, Alexander and Witten, Edward",
    title = "{Quantum Gravity Partition Functions in Three Dimensions}",
    eprint = "0712.0155",
    archivePrefix = "arXiv",
    primaryClass = "hep-th",
    doi = "10.1007/JHEP02(2010)029",
    journal = "JHEP",
    volume = "02",
    pages = "029",
    year = "2010"
}

@article{Regge:1974zd,
    author = "Regge, Tullio and Teitelboim, Claudio",
    title = "{Role of Surface Integrals in the Hamiltonian Formulation of General Relativity}",
    reportNumber = "Print-74-0988 (IAS,PRINCETON)",
    doi = "10.1016/0003-4916(74)90404-7",
    journal = "Annals Phys.",
    volume = "88",
    pages = "286",
    year = "1974"
}

@article{Szabados:2009eka,
    author = "Szabados, L{\'a}szl{\'o} B.",
    title = "{Quasi-Local Energy-Momentum and Angular Momentum in General Relativity}",
    doi = "10.12942/lrr-2009-4",
    journal = "Living Rev. Rel.",
    volume = "12",
    pages = "4",
    year = "2009"
}

@article{Booth:1998eh,
    author = "Booth, I. S. and Mann, Robert B.",
    title = "{Moving observers, nonorthogonal boundaries, and quasilocal energy}",
    eprint = "gr-qc/9810009",
    archivePrefix = "arXiv",
    doi = "10.1103/PhysRevD.59.064021",
    journal = "Phys. Rev. D",
    volume = "59",
    pages = "064021",
    year = "1999"
}

@article{Gibbons:1976ue,
    author = "Gibbons, G. W. and Hawking, S. W.",
    title = "{Action Integrals and Partition Functions in Quantum Gravity}",
    reportNumber = "PRINT-76-0995 (CAMBRIDGE)",
    doi = "10.1103/PhysRevD.15.2752",
    journal = "Phys. Rev. D",
    volume = "15",
    pages = "2752--2756",
    year = "1977"
}

@article{Brown:1992br,
    author = "Brown, J. David and York, Jr., James W.",
    title = "{Quasilocal energy and conserved charges derived from the gravitational action}",
    eprint = "gr-qc/9209012",
    archivePrefix = "arXiv",
    reportNumber = "IFP-423-UNC, TAR-009-UNC",
    doi = "10.1103/PhysRevD.47.1407",
    journal = "Phys. Rev. D",
    volume = "47",
    pages = "1407--1419",
    year = "1993"
}

@article{Dolan:2014mra,
    author = "Dolan, Brian P.",
    title = "{Vacuum energy and the latent heat of AdS-Kerr black holes}",
    eprint = "1407.4037",
    archivePrefix = "arXiv",
    primaryClass = "gr-qc",
    doi = "10.1103/PhysRevD.90.084002",
    journal = "Phys. Rev. D",
    volume = "90",
    number = "8",
    pages = "084002",
    year = "2014"
}

@article{Caldarelli:1999xj,
    author = "Caldarelli, Marco M. and Cognola, Guido and Klemm, Dietmar",
    title = "{Thermodynamics of Kerr-Newman-AdS black holes and conformal field theories}",
    eprint = "hep-th/9908022",
    archivePrefix = "arXiv",
    reportNumber = "UTF-434",
    doi = "10.1088/0264-9381/17/2/310",
    journal = "Class. Quant. Grav.",
    volume = "17",
    pages = "399--420",
    year = "2000"
}

@article{Chua:2023srl,
    author = "Chua, Wan Zhen and Hartman, Thomas",
    title = "{Black hole wavefunctions and microcanonical states}",
    eprint = "2309.05041",
    archivePrefix = "arXiv",
    primaryClass = "hep-th",
    doi = "10.1007/JHEP06(2024)054",
    journal = "JHEP",
    volume = "06",
    pages = "054",
    year = "2024"
}

@article{Marolf:2018ldl,
    author = "Marolf, Donald",
    title = "{Microcanonical Path Integrals and the Holography of small Black Hole Interiors}",
    eprint = "1808.00394",
    archivePrefix = "arXiv",
    primaryClass = "hep-th",
    doi = "10.1007/JHEP09(2018)114",
    journal = "JHEP",
    volume = "09",
    pages = "114",
    year = "2018"
}

@article{Compere:2008us,
    author = "Compere, Geoffrey and Marolf, Donald",
    title = "{Setting the boundary free in AdS/CFT}",
    eprint = "0805.1902",
    archivePrefix = "arXiv",
    primaryClass = "hep-th",
    doi = "10.1088/0264-9381/25/19/195014",
    journal = "Class. Quant. Grav.",
    volume = "25",
    pages = "195014",
    year = "2008"
}

@article{Brown:1992bq,
    author = "Brown, J. David and York, Jr., James W.",
    title = "{The Microcanonical functional integral. 1. The Gravitational field}",
    eprint = "gr-qc/9209014",
    archivePrefix = "arXiv",
    reportNumber = "IFP-441-UNC, TAR-028-UNC",
    doi = "10.1103/PhysRevD.47.1420",
    journal = "Phys. Rev. D",
    volume = "47",
    pages = "1420--1431",
    year = "1993"
}

@article{bams/1183524395,
    author = {Herman Gluck},
    title = {{The embedding of two-spheres in the four-sphere}},
    volume = {67},
    journal = {Bulletin of the American Mathematical Society},
    number = {6},
    publisher = {American Mathematical Society},
    pages = {586 -- 589},
    year = {1961},
}

@article{Maxfield:2020ale,
    author = "Maxfield, Henry and Turiaci, Gustavo J.",
    title = "{The path integral of 3D gravity near extremality; or, JT gravity with defects as a matrix integral}",
    eprint = "2006.11317",
    archivePrefix = "arXiv",
    primaryClass = "hep-th",
    doi = "10.1007/JHEP01(2021)118",
    journal = "JHEP",
    volume = "01",
    pages = "118",
    year = "2021"
}

@article{Mann:1996bi,
    author = "Mann, Robert B. and Solodukhin, Sergei N.",
    title = "{Conical geometry and quantum entropy of a charged Kerr black hole}",
    eprint = "hep-th/9604118",
    archivePrefix = "arXiv",
    reportNumber = "WATPHYS-TH-96-04",
    doi = "10.1103/PhysRevD.54.3932",
    journal = "Phys. Rev. D",
    volume = "54",
    pages = "3932--3940",
    year = "1996"
}

@article{Krishnan:2016mcj,
    author = "Krishnan, Chethan and Raju, Avinash",
    title = "{A Neumann Boundary Term for Gravity}",
    eprint = "1605.01603",
    archivePrefix = "arXiv",
    primaryClass = "hep-th",
    doi = "10.1142/S0217732317500778",
    journal = "Mod. Phys. Lett. A",
    volume = "32",
    number = "14",
    pages = "1750077",
    year = "2017"
}

@article{Krishnan:2016tqj,
    author = "Krishnan, Chethan and Kumar, K. V. Pavan and Raju, Avinash",
    title = "{An alternative path integral for quantum gravity}",
    eprint = "1609.04719",
    archivePrefix = "arXiv",
    primaryClass = "hep-th",
    doi = "10.1007/JHEP10(2016)043",
    journal = "JHEP",
    volume = "10",
    pages = "043",
    year = "2016"
}

@article{Hawking:1995ap,
    author = "Hawking, S. W. and Ross, Simon F.",
    title = "{Duality between electric and magnetic black holes}",
    eprint = "hep-th/9504019",
    archivePrefix = "arXiv",
    reportNumber = "DAMTP-R-95-8",
    doi = "10.1103/PhysRevD.52.5865",
    journal = "Phys. Rev. D",
    volume = "52",
    pages = "5865--5876",
    year = "1995"
}

@article{Blau:1991mp,
    author = "Blau, Matthias and Thompson, George",
    title = "{Quantum Yang-Mills theory on arbitrary surfaces}",
    reportNumber = "NIKHEF-H-91-09, MZ-TH-91-17",
    doi = "10.1142/S0217751X9200168X",
    journal = "Int. J. Mod. Phys. A",
    volume = "7",
    pages = "3781--3806",
    year = "1992"
}

@article{article,
author = {Blau, Matthias and Thompson, George},
year = {1991},
month = {05},
pages = {},
title = {Quantum Maxwell theory on arbitrary surfaces}
}

@article{Hristov:2021qsw,
    author = "Hristov, Kiril",
    title = "{4d $ \mathcal{N} $ = 2 supergravity observables from Nekrasov-like partition functions}",
    eprint = "2111.06903",
    archivePrefix = "arXiv",
    primaryClass = "hep-th",
    doi = "10.1007/JHEP02(2022)079",
    journal = "JHEP",
    volume = "02",
    pages = "079",
    year = "2022"
}

@article{Hristov:2022pmo,
    author = "Hristov, Kiril",
    title = "{The dark (BPS) side of thermodynamics in Minkowski$_{4}$}",
    eprint = "2207.12437",
    archivePrefix = "arXiv",
    primaryClass = "hep-th",
    doi = "10.1007/JHEP09(2022)204",
    journal = "JHEP",
    volume = "09",
    pages = "204",
    year = "2022"
}

@article{BenettiGenolini:2025jwe,
    author = "Benetti Genolini, Pietro and Murthy, Sameer",
    title = "{The gravitational index and allowable complex metrics}",
    eprint = "2503.20866",
    archivePrefix = "arXiv",
    primaryClass = "hep-th",
    doi = "10.1088/1751-8121/add7a7",
    journal = "J. Phys. A",
    volume = "58",
    number = "21",
    pages = "215401",
    year = "2025"
}

@article{Witten:2021nzp,
    author = "Witten, Edward",
    title = "{A Note On Complex Spacetime Metrics}",
    eprint = "2111.06514",
    archivePrefix = "arXiv",
    primaryClass = "hep-th",
    month = "11",
    year = "2021"
}

@article{Bandyopadhyay:2025jbc,
    author = "Bandyopadhyay, Subhodip and Punia, Gurmeet Singh and Srivastava, Yogesh K. and Virmani, Amitabh",
    title = "{The gravitational index of a small black ring}",
    eprint = "2504.09982",
    archivePrefix = "arXiv",
    primaryClass = "hep-th",
    month = "4",
    year = "2025"
}

@inbook{Cassani:2025sim,
    author = "Cassani, Davide and Murthy, Sameer",
    title = "{Quantum black holes: supersymmetry and exact results}",
    eprint = "2502.15360",
    archivePrefix = "arXiv",
    primaryClass = "hep-th",
    month = "2",
    year = "2025"
}

@article{Boruch:2025qdq,
    author = "Boruch, Jan and Emparan, Roberto and Iliesiu, Luca V. and Murthy, Sameer",
    title = "{The gravitational index of 5d black holes and black strings}",
    eprint = "2501.17909",
    archivePrefix = "arXiv",
    primaryClass = "hep-th",
    month = "1",
    year = "2025"
}

@article{Hegde:2023jmp,
    author = "Hegde, Subramanya and Virmani, Amitabh",
    title = "{Killing spinors for finite temperature Euclidean solutions at the BPS bound}",
    eprint = "2311.09427",
    archivePrefix = "arXiv",
    primaryClass = "hep-th",
    doi = "10.1007/JHEP02(2024)203",
    journal = "JHEP",
    volume = "02",
    pages = "203",
    year = "2024"
}

@article{BenettiGenolini:2023rkq,
    author = "Benetti Genolini, Pietro and Cabo-Bizet, Alejandro and Murthy, Sameer",
    title = "{Supersymmetric phases of AdS$_{4}$/CFT$_{3}$}",
    eprint = "2301.00763",
    archivePrefix = "arXiv",
    primaryClass = "hep-th",
    doi = "10.1007/JHEP06(2023)125",
    journal = "JHEP",
    volume = "06",
    pages = "125",
    year = "2023"
}

@article{Bobev:2020pjk,
    author = "Bobev, Nikolay and Charles, Anthony M. and Min, Vincent S.",
    title = "{Euclidean black saddles and AdS$_{4}$ black holes}",
    eprint = "2006.01148",
    archivePrefix = "arXiv",
    primaryClass = "hep-th",
    doi = "10.1007/JHEP10(2020)073",
    journal = "JHEP",
    volume = "10",
    pages = "073",
    year = "2020"
}

@article{Cabo-Bizet:2018ehj,
    author = "Cabo-Bizet, Alejandro and Cassani, Davide and Martelli, Dario and Murthy, Sameer",
    title = "{Microscopic origin of the Bekenstein-Hawking entropy of supersymmetric AdS$_{5}$ black holes}",
    eprint = "1810.11442",
    archivePrefix = "arXiv",
    primaryClass = "hep-th",
    doi = "10.1007/JHEP10(2019)062",
    journal = "JHEP",
    volume = "10",
    pages = "062",
    year = "2019"
}

@article{Iliesiu:2021are,
    author = "Iliesiu, Luca V. and Kologlu, Murat and Turiaci, Gustavo J.",
    title = "{Supersymmetric indices factorize}",
    eprint = "2107.09062",
    archivePrefix = "arXiv",
    primaryClass = "hep-th",
    month = "7",
    year = "2021"
}

@article{Heydeman:2020hhw,
    author = "Heydeman, Matthew and Iliesiu, Luca V. and Turiaci, Gustavo J. and Zhao, Wenli",
    title = "{The statistical mechanics of near-BPS black holes}",
    eprint = "2011.01953",
    archivePrefix = "arXiv",
    primaryClass = "hep-th",
    reportNumber = "PUPT-2621",
    doi = "10.1088/1751-8121/ac3be9",
    journal = "J. Phys. A",
    volume = "55",
    number = "1",
    pages = "014004",
    year = "2022"
}

@article{Cassani:2019mms,
    author = "Cassani, Davide and Papini, Lorenzo",
    title = "{The BPS limit of rotating AdS black hole thermodynamics}",
    eprint = "1906.10148",
    archivePrefix = "arXiv",
    primaryClass = "hep-th",
    doi = "10.1007/JHEP09(2019)079",
    journal = "JHEP",
    volume = "09",
    pages = "079",
    year = "2019"
}

@article{Bobev:2019zmz,
    author = "Bobev, Nikolay and Crichigno, P. Marcos",
    title = "{Universal spinning black holes and theories of class $ \mathcal{R} $}",
    eprint = "1909.05873",
    archivePrefix = "arXiv",
    primaryClass = "hep-th",
    doi = "10.1007/JHEP12(2019)054",
    journal = "JHEP",
    volume = "12",
    pages = "054",
    year = "2019"
}

@article{Larsen:2021wnu,
    author = "Larsen, Finn and Lee, Siyul",
    title = "{Microscopic entropy of AdS$_{3}$ black holes revisited}",
    eprint = "2101.08497",
    archivePrefix = "arXiv",
    primaryClass = "hep-th",
    reportNumber = "LCTP-21-03",
    doi = "10.1007/JHEP07(2021)038",
    journal = "JHEP",
    volume = "07",
    pages = "038",
    year = "2021"
}

@article{Anupam:2023yns,
    author = "Anupam, A. H. and Chowdhury, Chandramouli and Sen, Ashoke",
    title = "{Revisiting Logarithmic Correction to Five Dimensional BPS Black Hole Entropy}",
    eprint = "2308.00038",
    archivePrefix = "arXiv",
    primaryClass = "hep-th",
    month = "7",
    year = "2023"
}

@article{H:2023qko,
    author = "H., Anupam A. and Athira, P. V. and Chowdhury, Chandramouli and Sen, Ashoke",
    title = "{Logarithmic Correction to BPS Black Hole Entropy from Supersymmetric Index at Finite Temperature}",
    eprint = "2306.07322",
    archivePrefix = "arXiv",
    primaryClass = "hep-th",
    month = "6",
    year = "2023"
}

@article{Hegde:2024bmb,
    author = "Hegde, Subramanya and Sen, Ashoke and Shanmugapriya, P. and Virmani, Amitabh",
    title = "{Supersymmetric Index for Half BPS Black Holes in N=2 Supergravity with Higher Curvature Corrections}",
    eprint = "2411.08260",
    archivePrefix = "arXiv",
    primaryClass = "hep-th",
    month = "11",
    year = "2024"
}

@article{Chowdhury:2024ngg,
    author = "Chowdhury, Chandramouli and Sen, Ashoke and Shanmugapriya, P. and Virmani, Amitabh",
    title = "{Supersymmetric index for small black holes}",
    eprint = "2401.13730",
    archivePrefix = "arXiv",
    primaryClass = "hep-th",
    doi = "10.1007/JHEP04(2024)136",
    journal = "JHEP",
    volume = "04",
    pages = "136",
    year = "2024"
}

@article{Boruch:2023gfn,
    author = "Boruch, Jan and Iliesiu, Luca V. and Murthy, Sameer and Turiaci, Gustavo J.",
    title = "{New forms of attraction: Attractor saddles for the black hole index}",
    eprint = "2310.07763",
    archivePrefix = "arXiv",
    primaryClass = "hep-th",
    month = "10",
    year = "2023"
}

@article{Heydeman:2024fgk,
    author = "Heydeman, Matthew and Toldo, Chiara",
    title = "{Mixed 't Hooft Anomalies and the Witten Effect for AdS Black Holes}",
    eprint = "2412.03695",
    archivePrefix = "arXiv",
    primaryClass = "hep-th",
    month = "12",
    year = "2024"
}

@article{Adhikari:2024zif,
    author = "Adhikari, Soumya and Dharanipragada, Pavan and Goswami, Kaberi and Virmani, Amitabh",
    title = "{Attractor saddle for 5D black hole index}",
    eprint = "2411.12413",
    archivePrefix = "arXiv",
    primaryClass = "hep-th",
    month = "11",
    year = "2024"
}

@article{Cassani:2024kjn,
    author = "Cassani, Davide and Ruip\'erez, Alejandro and Turetta, Enrico",
    title = "{Localization of the 5D supergravity action and Euclidean saddles for the black hole index}",
    eprint = "2409.01332",
    archivePrefix = "arXiv",
    primaryClass = "hep-th",
    month = "9",
    year = "2024"
}

@article{Chen:2024gmc,
    author = "Chen, Yiming and Murthy, Sameer and Turiaci, Gustavo J.",
    title = "{Gravitational index of the heterotic string}",
    eprint = "2402.03297",
    archivePrefix = "arXiv",
    primaryClass = "hep-th",
    doi = "10.1007/JHEP09(2024)041",
    journal = "JHEP",
    volume = "09",
    pages = "041",
    year = "2024"
}

@article{Braden:1990hw,
      author         = "Braden, Harry W. and Brown, J. David and Whiting, Bernard
                        F. and York, Jr., James W.",
      title          = "{Charged black hole in a grand canonical ensemble}",
      journal        = "Phys. Rev.",
      volume         = "D42",
      year           = "1990",
      pages          = "3376-3385",
      doi            = "10.1103/PhysRevD.42.3376",
      reportNumber   = "PRINT-90-0461 (NORTH-CAROLINA)",
      SLACcitation   = "%%CITATION = PHRVA,D42,3376;%%"
}

@article{Iliesiu:2020qvm,
    author = "Iliesiu, Luca V. and Turiaci, Gustavo J.",
    title = "{The statistical mechanics of near-extremal black holes}",
    eprint = "2003.02860",
    archivePrefix = "arXiv",
    primaryClass = "hep-th",
    month = "3",
    year = "2020"
}

@article{Cordes:1994fc,
      author         = "Cordes, Stefan and Moore, Gregory W. and Ramgoolam,
                        Sanjaye",
      title          = "{Lectures on 2-d Yang-Mills theory, equivariant
                        cohomology and topological field theories}",
      booktitle      = "{NATO Advanced Study Institute: Les Houches Summer
                        School, Session 62: Fluctuating Geometries in Statistical
                        Mechanics and Field Theory Les Houches, France, August
                        2-September 9, 1994}",
      journal        = "Nucl. Phys. Proc. Suppl.",
      volume         = "41",
      year           = "1995",
      pages          = "184-244",
      doi            = "10.1016/0920-5632(95)00434-B",
      eprint         = "hep-th/9411210",
      archivePrefix  = "arXiv",
      primaryClass   = "hep-th",
      reportNumber   = "YCTP-P11-94",
      SLACcitation   = "%%CITATION = HEP-TH/9411210;%%"
}
\end{document}